\documentclass[useAMS, usenatbib]{mn2e}

\usepackage{graphicx}
\usepackage{amssymb,amsmath}
\usepackage{natbib}

\newcommand{\beq}	{\begin{equation}}
\newcommand{\eeq}	{\end{equation}}
\newcommand{\beqa}	{\begin{eqnarray}}
\newcommand{\eeqa}	{\end{eqnarray}}

\newcommand{\gad}       {{\gamma_{\rm AD}}}

\newcommand{\alfven}    {{Alfv$\acute{\rm e}$n }}

\newcommand{\alfvenic}  {{Alfv$\acute{\rm e}$nic }}

\newcommand{\e}	        {$^{-1}$}

\newcommand{\eee}	{$^{-3}$}

\newcommand{\nbh}	{\bar n_{\rm H}}

\newcommand{\radlo}	{R_{\rm AD}(\ell_0)}

\usepackage{rotating}
\usepackage[draft]{hyperref}

\title[The CH$^+$ Abundance in Turbulent, Diffuse Molecular Clouds]{The CH$^+$ Abundance in Turbulent, Diffuse Molecular Clouds}

\author[Myers, McKee, \& Li]
        {Andrew T. Myers$^1$,
        Christopher F. McKee$^{2,3}$, and
        Pak Shing Li$^3$ \\
        $^1$ Lawrence Berkeley National Laboratory, 
        	        1 Cyclotron Road, Berkeley, CA, 94720, USA; atmyers@lbl.gov \\
        $^2$  Department of Physics, University of California, Berkeley,
                   Berkeley, CA 94720 \\
         $^3$ Department of Astronomy, University of California, Berkeley,
                   Berkeley, CA 94720 \\}

\begin{document}

\label{firstpage}

\maketitle

\begin{centering}
\section*{Abstract}
\end{centering}

The intermittent dissipation of interstellar turbulence is an important energy source in the diffuse ISM. Though on average smaller than the heating rates due to cosmic rays and the photoelectric effect on dust grains, the turbulent cascade can channel large amounts of energy into a relatively small fraction of the gas that consequently undergoes significant heating and chemical enrichment. In particular, this mechanism has been proposed as a solution to the long-standing problem of the high abundance of CH$^+$ along diffuse molecular sight lines, which steady-state, low temperature models under-produce by over an order of magnitude. While much work has been done on the structure and chemistry of these small-scale dissipation zones, comparatively little attention has been paid to relating these zones to the properties of the large-scale turbulence. In this paper, we attempt to bridge this gap by estimating the temperature and CH$^+$ column density along diffuse molecular sight-lines by post-processing 3-dimensional MHD turbulence simulations. Assuming reasonable values for the cloud density ($\bar{n}_{\rm{H}} = 30 $ cm$^{-3}$), size ($L =$ 20 pc), and velocity dispersion ($\sigma_v = 2.3$ km s$^{-1}$), we find that our computed abundances compare well with CH$^+$ column density observations, as well as with observations of emission lines from rotationally excited H$_2$ molecules.

\section{Introduction}

The CH$^+$ ion is commonly detected along sight lines towards bright O and B stars, with column densities $\gtrsim 10^{13}$ cm$^{-2}$ frequently reported in the literature \citep[e.g.][]{gredel1993, gre1997, cra1995, wes2008, she2008}. This prevalence is puzzling, however, because CH$^+$ is destroyed very efficiently by both atomic and molecular hydrogen, and the only reaction that can form CH$^+$ rapidly,
\beq
\label{reaction} 
\rm{C}^+ + \rm{H}_2 \rightarrow \rm{CH}^+ + {\rm{H}} \hspace{0.3 in}\Delta E / k = -4640 \,\rm{ K},
\eeq 
is strongly endothermic and can only proceed at temperatures of $ \sim 1000$ K or higher. For this reason, models of diffuse interstellar clouds with $T \lesssim 100$ K, like those of \cite{vanD1986}, fail dramatically to reproduce these high CH$^+$ columns, despite their success with other species. 

Most proposed solutions to this problem have invoked an additional energy source to overcome this 4640 K activation barrier. Possibilities include hydrodynamic \citep{elitzur1978, elitzur1980} and magnetohydrodynamic \citep{draine86} shock waves,  heating in turbulent boundary layers at cloud surfaces \citep{duley1992}, and particularly dense photon-dominated regions (PDRs) surrounding bright stars \citep{duley1992, sternberg1995}; for an overview of these mechanisms and some of the problems they face confronting observations, see \cite{gre1997}. A particularly promising idea, pioneered by \cite{falgaronepuget1995}, is that the intermittent dissipation of turbulence heats small regions within diffuse clouds to the $\gtrsim 1000$ K temperatures required for (1) to proceed. Drawing on laboratory experiments of unmagnetized, incompressible turbulent flows, they calculated that if the velocity dispersion in cold, mostly atomic clouds at a scale of 1 pc is 3 km s$^{-1}$, then a few percent of the cloud could be heated to $> 1000$ K, a mass fraction sufficient to bring the CH$^+$ abundance in line with observed values \citep{lambert86}. This result was later found to be consistent with magnetized, compressible turbulence simulations as well \citep{pan2009}. These pockets of warm gas may also explain the observed emission from the first few excited rotational states of H$_2$ detected in diffuse gas, which is often too large to be explained by UV pumping alone \citep[e.g.][]{falg2005, goldsmith2010, ingalls2011}. 

Models that rely on turbulent heating alone can over-predict the abundance of other species, such as OH, which is already well modeled by cold cloud models \citep{fed1996}. However, in addition to the direct heating effect, turbulence can give rise to net drift velocities between the ionic and neutral species in plasmas, enhancing the rates of ion-neutral reactions like (1) beyond those expected from the kinetic temperature alone \citep[e.g.][]{dra1980, flower1985}. \cite{fed1996} approximated this effect by computing the rate of reaction (1) at the effective temperature $T_{\rm eff}$ given by:
\beq
\label{eq:Teff}
T_{\rm{eff}} = T + \frac{\mu}{3 k} v_d^2,
\eeq where $\mu$ is the reduced mass of (1) and  $v_d$ is the magnitude of the ion-neutral drift velocity. They proposed that MHD waves with amplitudes $\sim 3$ km s$^{-1}$ can enhance the predicted column densities of CH$^+$ to the observed values even in gas that remains $T \lesssim 100$ K. A similar calculation was made in \cite{spaans1995}, who computed the distribution of $v_d$ from an analytic intermittency model. More recently, \cite{she2008} included this effect in their PDR models, finding a similar result. The appeal of these models is that they have fewer problems over-producing molecules such as OH, which is not formed by an ion-neutral reaction.   

The most successful models include both of these effects simultaneously. \cite{jou1998} and \cite{god2009} treat regions of intense dissipation, termed ``Turbulent Dissipation Regions" (TDRs) by \citeauthor{god2009}, as magnetized vortices, taking their (axisymmetric) velocity profiles from that of a Burgers vortex, for which the vorticity as a function of radius is 
\beq
\omega(r) = \omega_0 \exp{\left[-\left(\frac{r}{r_0}\right)^2 \right]}.
\eeq
Here, $\omega_0$ and $r_0$ are parameters describing the peak vorticity and characteristic fall-off radius in the vortex. Typically, $\omega_0 \approx 6 \times 10^{-10}$ s$^{-1}$ and $r_0 \approx 40$ AU in \cite{god2009}. These calculations follow the subsequent thermal and chemical evolution of parcels of gas trapped inside such a vortex, including both turbulent heating and ion-neutral drift. \cite{god2009} then construct models of entire sight lines by assuming they intersect some number of these vortex structures to account for the observed column density of CH$^+$. These models have had a great deal of success reproducing the observed CH$^+$ and excited H$_2$ columns without overproducing species such as OH. 

The goal of this paper is to provide a complementary approach to the above models, which concentrate on individual dissipation events. We post-process the gas temperature $T$, drift velocity $v_d$, and CH$^+$ abundance cell-by-cell through an output of a turbulence simulation from \cite{li2012b} that has been scaled to typical diffuse cloud conditions.  This approach loses some of the detail of the above models, but it has the advantage of making fewer simplifying assumptions about, for example, the nature of the intermittent structures or the number of dissipation events along a line of sight. We find that CH$^+$ columns in excess of $\sim 10^{13}$ cm$^{-2}$ are readily obtained. We compare our results against a statistically homogenous sample of CH$^+$-containing sight lines from \cite{wes2008} and against observations of rotationally excited H$_2$, finding good agreement with both.    

\section{Methodology}

\begin{table}
\begin{center}
\caption{ \label{tbl-1} Standard Physical and Chemical Model Parameters}
\begin{tabular}{lc}
\\
\hline \hline
$\bar{n}_{\rm{H}}$     &30 cm$^{-3}$ \\
$L$  & 20 pc \\
$\sigma_{\rm 1D} $  &2.3 km s$^{-1}$ \\
$\bar T_{\rm M} $ & 65 K \\
$T_{50, \rm{ M}}$ & 35 K \\
$B_{\rm{rms}}$ & 5.2 $\mu$G \\
$x(\rm{H})$  & $0.68$ \\
$x(\rm{H}_2)$  & $0.16$ \\
$x(\rm{He})$  & $0.1$ \\
$x(e^-)$ & $1.6\times 10^{-4} $ \\
$x(\rm{C})$  & $1.6\times 10^{-4} $ \\
$x(\rm{O})$  & $3.2\times 10^{-4} $ \\
\hline
\end{tabular}
\end{center}
\end{table}

In this section, we provide an overview of our calculation, including our treatment of the heating and cooling rates, the drift velocity, and our calculation of the CH$^+$ abundance. 

\subsection{Model Description}

\begin{figure*}
  \centering
    \includegraphics[width=1.0\textwidth]{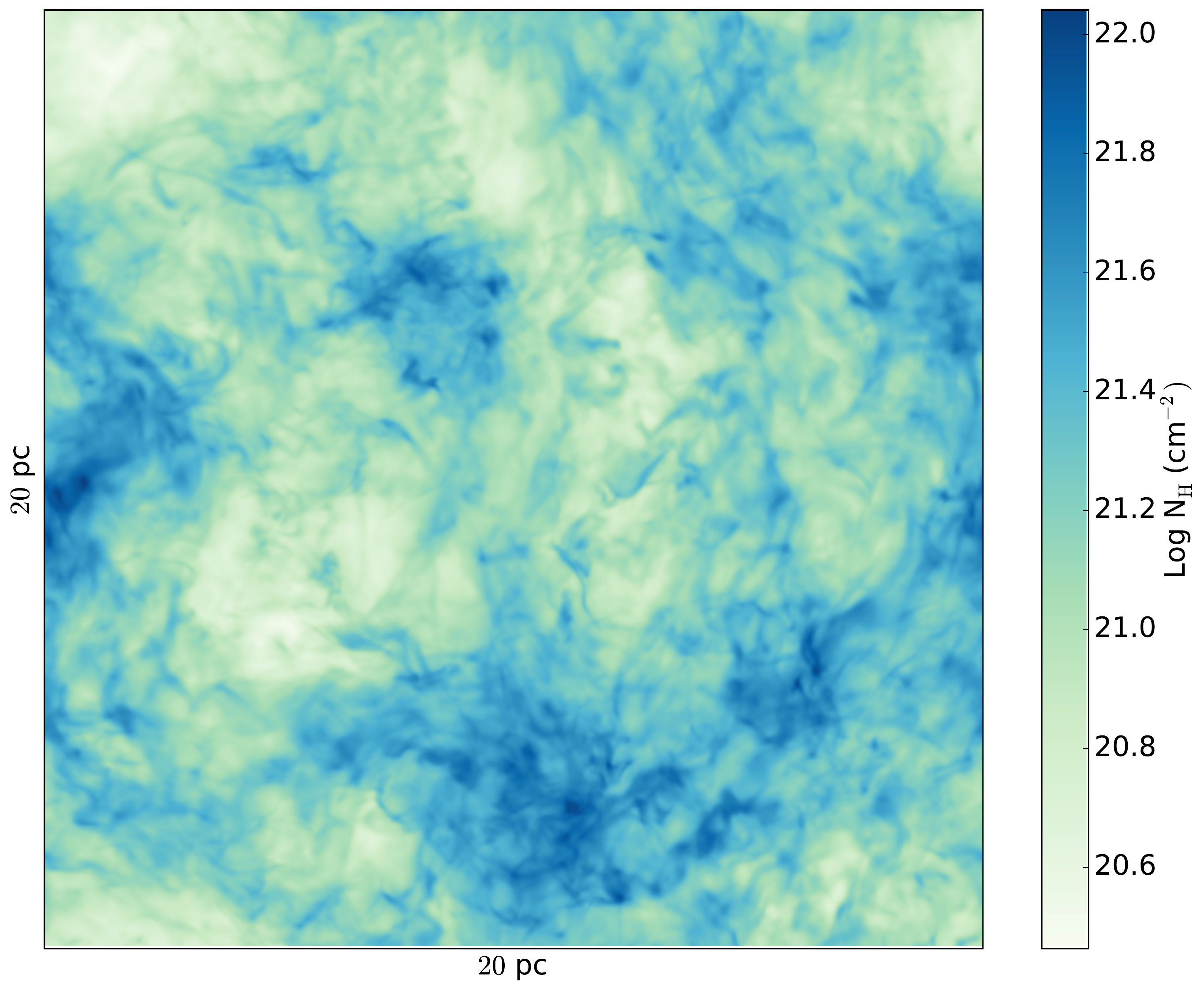}
    \caption{\label{column_density} Logarithm of the total column density $N_{\rm{H}}$ through the computational domain along the $x$ direction. The simulation data has been scaled such that the mean $N_{\rm{H}}$ is $1.83 \times 10^{21}$ cm$^{-2}$. The size of the box is indicated on the $x$ and $y$ axes.}
\end{figure*}

The CH$^+$ ion is believed to form in partially molecular environments. Indeed, in order for reaction (1) to proceed, at least some of the hydrogen must be in the form of H$_2$, and at least some of the carbon must be in C$^+$. Any plausible formation mechanism is thus not likely to be effective in either the outskirts of molecular clouds with visual extinction $A_V < 0.1$ mag, where the hydrogen is almost all atomic, or deep in their interiors at $A_V$ greater than a few, where almost all the carbon will be C and/or CO. This gas is sometimes referred to as the ``dark gas" since it is difficult to observe \citep{grenier2005, wol2010}.

We therefore model interstellar clouds in which the hydrogen has begun to turn to H$_2$, but the carbon is still primarily in the form of C$^+$. \cite{snow2006} classify such clouds as ``diffuse molecular clouds." We treat these regions as cubic boxes with length $\ell_0$, mean hydrogen nucleus number density $\bar{n}_{\rm H}$, and one-dimensional velocity dispersion $\sigma_{\rm 1D}$. The mean mass per hydrogen nucleus is $\mu_{\rm{H}} = 2.34 \times 10^{-24}$ g cm$^{-3}$. For simplicity, we set the relative abundances (relative to hydrogen nuclei) of molecular hydrogen $x(\rm{H}_2)$ and ionized carbon $x(\rm{C}^+)$ to be constant across the region. For the former, we adopt $x(\rm{H}_2) = 0.16$, the mean observed molecular fraction from the sample of \cite{wes2008}, which studied the correlation of CH$^+$ column density with that of atomic and molecular hydrogen. For the latter, we take $x(\rm{C}^+) = 1.6 \times 10^{-4}$ from \cite{sof2004}. \cite{sofia2011} find using a different measurement technique that the gas-phase carbon abundance is lower than that adopted here by a factor of $\approx 0.43$. Since it is not clear which measurement is more accurate, we choose to adopt the higher value. The effects of a lower C abundance in our model are complex. On one hand, it directly reduces the CH$^+$ formation rate (see Section 2.5), since C$^+$ is one of the reactants in (1). On the other hand, it decreases the cooling rate due to C$^+$ (Section 2.3) and increases our estimate for the ion-neutral drift velocity (Section 2.4). The overall effect of deceasing our assumed carbon abundance by a factor of $0.43$ is to \emph{increase} the estimate for the CH$^+$ abundance by approximately 30\%.

Although chemical and physical models of diffuse gas often assume a constant $n_{\rm{H}}$, the density distribution in the ISM in fact contains a wide range of fluctuations over many orders of magnitude due to the compressive effects of supersonic turbulence. To treat this, we use the results of a 512$^3$ driven, turbulence simulation first published in \cite{li2012b}. The density, magnetic field, and velocity at every point in our model are drawn from the corresponding cell in a data dump from this simulation, scaled to physical units by the process described below. A color plot of the column density through the simulation volume is shown in Figure \ref{column_density}. An important caveat to our calculation is that this simulation data is \emph{isothermal}. We then calculate what the temperature would be if the intermittency in the isothermal case were the same as if the time-dependent heating and cooling effects were followed self-consistently.  

\subsection{Scaling to Physical Units}

Simulations of magnetized, isothermal turbulent boxes are characterized by two dimensionless numbers: the 3D sonic Mach number $\mathcal{M} = \sigma_{\rm{3D}} / c_s$ and the 3D \alfvenic Mach number  $\mathcal{M}_{\rm{A}} =  \sigma_{\rm{3D}} / v_{\rm{A}}$. Here, $\sigma_{\rm{3 D}}$ is the three-dimensional, density-weighted, non-thermal velocity dispersion in the box, $c_s = \sqrt{k T / \bar m}$ is the isothermal sound speed, where $T$ is the temperature and $\bar m$ the mean mass per particle, and $v_{\rm{A}} = B_{\rm{rms}} / \sqrt{4 \pi \bar \rho}$ is the \alfven velocity, where $B_{\rm{rms}}$ is the root-mean-square magnetic field and $\bar \rho$ the mean density. In the simulation considered here, the turbulence was driven so as to maintain $\mathcal{M} \approx 10$, and the initial $\mathcal{M}_{\rm{A}}$ was $\sqrt{5}$. 

In the absence of further constraints, we would be free to scale $\bar \rho$, $\sigma_{\rm{3D}}$, $c_s$, $B_{\rm{rms}}$ and the size of the box $\ell_0$ at will as long as the dimensionless ratios $\mathcal{M}$ and $\mathcal{M}_{\rm{A}}$ remained invariant (see \cite{mckee2010} for a more rigorous discussion of scaling laws for turbulent box simulations). However, to be consistent with observations of diffuse molecular gas, we impose several additional constraints. First, we require that the gas in the box obey a linewidth-size relation \citep[e.g.][]{mckee2007}:
\beq
\label{eq:ls}
\sigma_{\rm{1D}} = \sigma_{\rm pc} R_{\rm{pc}}^{0.5},
\eeq 
where $R_{\rm{pc}}$ is the cloud radius in parsecs, and $\sigma_{\rm pc} = 0.72$ km s$^{-1}$. The 1D non-thermal velocity dispersion $\sigma_{\rm{1D}}$ is related to the 3D value by $\sigma_{\rm{3D}} = \sqrt{3} \sigma_{\rm{1D}}$, and in applying Equation (\ref{eq:ls}), which is meant for approximately spherical clouds, to our cubic simulation domain, we identify the cloud radius $R$ with $\ell_0 / 2$. Second, we require that the mean column density of hydrogen nuclei be fixed:
\beq
\label{eq:column}
\bar N_{\rm{H}} = \bar{n}_{\rm{H}} \ell_0 = \bar N_{\rm{obs}},
\eeq 
where $\bar N_{\rm{obs}} \approx 1.83 \times 10^{21}$ cm$^{-2}$ is the mean total column density from \cite{wes2008}. This column corresponds to $A_V \approx 1$, consistent with our requirement that the gas be partially molecular. Note that Equations (\ref{eq:ls}) and (\ref{eq:column}) imply that we cannot independently choose $n_{\rm{H}}$, $\ell_0$ and $\sigma_{\rm{1D}}$; choosing a density fixes the box size $\ell_0$, which fixes the velocity dispersion through the linewidth-size relation. Numerically:
\begin{align}
\ell_0 \approx & \hspace{3 pt} 19.8 \hspace{3 pt} \left( \frac{30 \hspace{3 pt} \rm{cm}^{-3}}{\bar n_{\rm{H}}} \right) \hspace{3 pt} \rm{pc}, \nonumber \\
\sigma_{\rm{1D}} \approx & \hspace{3 pt} 2.3 \hspace{3 pt} \left( \frac{30 \hspace{3 pt} \rm{cm}^{-3}}{\bar n_{\rm{H}}} \right)^{0.5} \hspace{3 pt} \rm{km} \hspace{3 pt} \rm{s}^{-1}\ .
\end{align} The dynamical timescale in our model can thus be estimated as:
\beq
\label{eq:dynamical_time}
t_{\rm{dyn}} = \frac{\ell_0}{\sigma_{1D}} \approx 8.5 \times 10^6 \left( \frac{30 \hspace{3 pt} \rm{cm}^{-3}}{\bar n_{\rm{H}}} \right)^{0.5} \rm{yr}.
\eeq We assume that properties of the fluid flow (the density, velocity, and magnetic field) change on this timescale. We show that this is large compared to the thermal and chemical timescales below.

These relations also fix (along with the fact the $\mathcal{M}_{\rm{A}} = 2.2$) the rms magnetic field strength in the box:
\beq
B_{\rm{rms}} = \sqrt{\frac{6 \pi \mu_H \bar N_{\rm{obs}} \sigma_{\rm{pc}}^2}{ (1 \hspace{3 pt} \rm{pc}) \hspace{3 pt} \mathcal{M}_{\rm{A}}^2}} \approx 5.3 \hspace{3 pt} \mu \rm{G}.
\eeq 
\citet{cru2010} infer that interstellar fields in gas with $\bar n_{\rm H}\la 300$ cm\eee\ are uniformly distributed in strength between very low values and $10\,\mu$G, so this field is quite typical. Note that
only one of $\bar{n}_{\rm{H}}$, $\ell_0$, $\sigma_{\rm{3D}}$, and $B_{\rm{rms}}$ may be set independently, with the others following from that choice.   

The final remaining dimensional parameter describing our turbulence simulation is the isothermal sound speed $c_s$. The sound speed of a gas with $x(\rm{H}_2) = 0.16$ and $x(\rm{He}) = 0.1$ is
\beq
c_s(T) \approx 0.74 \hspace{3 pt} \sqrt{\frac{T}{100 \hspace{3 pt} \rm{K}}} \hspace{7 pt} \rm{km} \hspace{3 pt} \rm{s}^{-1}. 
\eeq  
However, because the temperature is an output of our model, we cannot set it arbitrarily. We thus compute the temperature using the process described below for a range of boxes, each scaled to a different $\bar n_{\rm{H}}$, and select the one for which $\mathcal{M}$ computed using the mass-weighted median temperature $T_{50, \rm{M}}$ (the $T$ for which half of the mass in cloud is hotter) is $\approx 10$. We find that this occurs at $n_{\rm{H}} \approx 30$ cm$^{-3}$ and adopt that as our fiducial density. This is the same value adopted in the standard model of \cite{jou1998}. However, individual sight lines passing through the simulation volume can have mean densities ranging from $\approx 5$ cm$^{-3}$ to  $\approx 180$ cm$^{-3}$. The corresponding $T_{50, \rm{M}}$ is $\approx 35$ K, and the mass-weighted mean temperature is $\bar T_{\rm M} \approx 65$ K. We summarize the physical and chemical parameters describing this model in Table \ref{tbl-1}.  

Our simulation data does not include the effects of self-gravity. As a consistency check, we compute the virial parameter $\alpha_{\rm{vir}}$ to verify that turbulence indeed dominates self-gravity. Using the above parameters, the total mass $M$ in the box is $\approx 8000$  $M_{\odot}$. The virial parameter for a spherical cloud of mass $M$ and radius $R$ is $\alpha_{\rm{vir}} = 5 \sigma_{\rm{1D}}^2 R / G M$ \citep{bertoldi1992}. Using $R = \ell_0/2$, we find $\alpha_{\rm{vir}} \approx 7.4$, which is large enough that the effects of self-gravity are indeed negligible. We can also characterize the relative importance of gravity and magnetic fields in the box by computing the mass-to-flux ratio relative to critical, $\mu_{\Phi} \equiv M / M_{\Phi}$, where $M_{\Phi} \approx \Phi / 2 \pi \sqrt{G}$ and $\Phi$ is the magnetic flux threading the cloud. For our adopted parameters, $\mu_{\Phi} \approx 1.3$, so the cloud is marginally magnetically supercritical. 

\subsection{Heating and Cooling}

\begin{figure}
  \centering
    \includegraphics[width=0.45\textwidth]{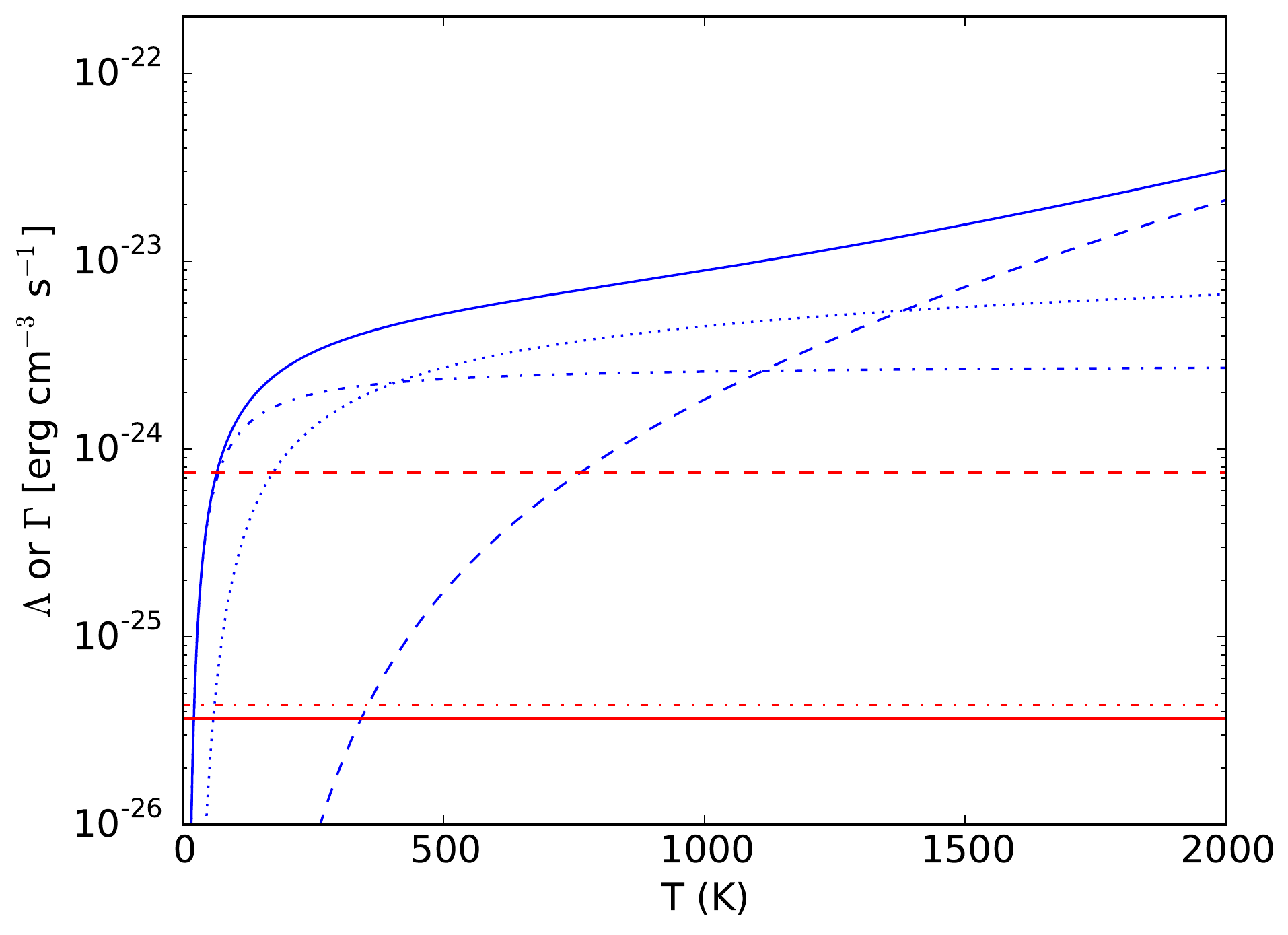}
    \caption{ \label{fig:energy_balance} Heating (red) and cooling (blue) rates per unit volume versus $T$ for $n_{\rm{H}} = 30$ cm$^{-3}$ and the standard chemical abundances shown in Table \ref{tbl-1}. The solid blue curve shows $ n_{\rm{H}} \Lambda_{\rm{tot}}$, while the dashed, dashed-dotted, and dotted curves are $ n_{\rm{H}} \Lambda_{\rm{H}_2}$, $ n_{\rm{H}} \Lambda_{\rm{C}^+}$, and $ n_{\rm{H}} \Lambda_{\rm{O}}$, respectively. The solid, dashed, and dashed-dotted red lines show the mean values of $\Gamma_{\rm{Turb}}$, $\Gamma_{\rm{PE}}$, and $\Gamma_{\rm{CR}}$, respectively.}
\end{figure}

The temperature in each cell is set by a balance between heating and cooling: 
\beq
\label{eq:ebalance}
\Gamma_{\rm{Turb}} + \Gamma_{\rm{CR}} + \Gamma_{\rm{PE}} = n_{\rm H} \Lambda_{\rm{tot}}(T).
\eeq where $\Gamma_{\rm{Turb}}$,  $\Gamma_{\rm{CR}}$, and $\Gamma_{\rm{PE}}$ are the heating rates per unit volume due to the dissipation of turbulence, cosmic ray ionizations, and the photoelectric effect on dust grains, respectively. $n_{\rm H} \Lambda_{\rm{tot}}(T)$ is the total cooling rate per unit volume, which we assume is dominated by electronic transitions of C$^+$ and O and by ro-vibrational transitions of the H$_2$ molecule. Note that throughout this paper, we use $\lambda$ to represent the cooling rate \emph{coefficient} (erg cm$^3$ s$^{-1}$) and $\Lambda$ for the cooling rate per H nucleus (erg s$^{-1}$). 

$\Gamma_{\rm{CR}}$ can be expressed as the product of three factors - the total cosmic ray ionization rate per H nucleus $\zeta_{\rm{H}}$ (including both primary and secondary ionizations), the average energy deposited into the medium per ionization $\Delta Q$, and $n_{\rm H}$. Both $\zeta_{\rm{H}}$ and $\Delta Q$ are rather uncertain and can vary considerably over different Galactic environments. Typical values of $\zeta_{\rm{H}}$ in dense gas are $\sim 1 - 5 \times 10^{-17}$ s$^{-1}$  \citep{dalgarno2006}, but there is evidence from ${\rm{H}_3}^+$ observations that $\zeta_{\rm{H}}$ is considerably higher in the diffuse gas under consideration here \citep{dalgarno2006, indriolo2012}. \cite{indriolo2012} find a mean $\zeta_{\rm{H}}$ of $1.8 \times 10^{-16}$ s$^{-1}$ in their sample of diffuse molecular sight lines, and values as large as  $\sim 1 \times 10^{-15}$ s$^{-1}$ have been reported in the literature \citep{snow2006, shaw2008}. In this paper, we adopt the value $1.8 \times 10^{-16}$ s$^{-1}$. For $\Delta Q$, we use 10 eV, as estimated for diffuse molecular gas from Table 6 of \cite{glassgold2012}, although it is important to note that this value can vary by several eV depending on the precise physical and chemical conditions in the cloud. Combining these factors, the cosmic ray heating rate is:

\begin{align}
\Gamma_{\rm{CR}} &= \zeta \Delta Q n_{\rm{H}} \nonumber \\
&\approx 1.9 \times 10^{-25} \left( \frac{n_{\rm{H}}}{30 \hspace{3 pt} \rm{cm}^{-3}} \right) \hspace{3 pt} \rm{ergs} \hspace{3 pt} \rm{cm}^{-3} \hspace{3 pt} \rm{s}^{-1}
\end{align} 

For $\Gamma_{\rm{PE}}$, we adopt the expression:
\beq
\Gamma_{\rm{PE}} = 1.3 \times 10^{-24} \hspace{3 pt} n_{\rm{H}} \epsilon G_0 \hspace{3 pt} \rm{ergs} \hspace{3 pt} \rm{cm}^{-3} \hspace{3 pt} \rm{s}^{-1}
\eeq from \cite{wol2003}, where $G_0$ is the intensity of FUV light in units of the \cite{habing1968} field and $\epsilon$ is a heating efficiency factor given by Equation (20) of \cite{wol2003}. For $n_{\rm{H}} = 30$ cm$^{-3}$, $T = 100$ K, an electron fraction of $1.6 \times 10^{-4}$, and a FUV field of $G_0 = 1.1$ \citep{mathis1983}, $\epsilon$ evaluates to $1.8 \times 10^{-2}$, yielding
\beq
\Gamma_{\rm{PE}} = 7.6 \times 10^{-25} \left( \frac{n_{\rm{H}}}{30 \hspace{3 pt} \rm{cm}^{-3}} \right) \hspace{3 pt} \rm{ergs} \hspace{3 pt} \rm{cm}^{-3} \hspace{3 pt} \rm{s}^{-1}
\eeq

The final heating process we consider is $\Gamma_{\rm{Turb}}$. Dimensional arguments \citep[e.g.][]{landau1959} and numerical simulations \citep[e.g.][]{stone1998, maclow1999} both suggest that the kinetic energy in a turbulent cloud $1/2 \rho \sigma_{3 \rm{D}}^2$ decays in roughly one crossing time $\ell_0 / \sigma_{3 \rm{D}}$, so that the volume-averaged turbulent heating rate is approximately

\begin{align}
\label{eq:av_turb}
\bar{\Gamma}_{\rm{Turb}}  &\approx \frac{1}{2} \frac{\rho \sigma_{3 \rm{D}}^3}{\ell_0} \nonumber \\
&\approx 3.5 \times 10^{-26} \sqrt{\frac{n_{\rm{H}}}{30 \hspace{3 pt} \rm{cm}^{-3}}} \hspace{3 pt} \rm{ergs} \hspace{3 pt} \rm{cm}^{-3} \hspace{3 pt} \rm{s}^{-1}, 
\end{align} where in the last step we have assumed the scaling given by Equations (\ref{eq:ls}) and (\ref{eq:column}). Locally, however, $\Gamma_{\rm{Turb}}$ exhibits large fluctuations from place to place, a phenomenon known as intermittency.  To calculate the spatial dependence of $\Gamma_{\rm{Turb}}$, we follow the calculation in \cite{pan2009}. To summarize their argument, the work done against the viscous forces in a fluid with velocity field $\bold v$ is irreversibly converted into heat at rate per unit volume given by: 
\beq
\label{eq:dissipation}
\Gamma_{\rm{Turb}}(\bold x) = \frac{1}{2} \rho \nu \left( \partial_i v_j + \partial_j v_i - \frac{2}{3} \left( \nabla \cdot \bold v \right) \delta_{ij} \right)^2.
\eeq This rate depends on the kinematic viscosity of the fluid, $\nu$. However, in our simulations, which were based on the Euler equations for a compressible gas, the viscosity was numerical in origin, and thus does not have its true microphysical value. Instead, we treat $\nu$ as a proportionality constant that takes whatever value is required so that the volume average of Equation (\ref{eq:dissipation}) equals Equation (\ref{eq:av_turb}). Once this constant has been determined, we can compute $\Gamma_{\rm{Turb}}(\bold x)$ for each cell in the simulation.

Note that the resolution of our simulation data $\Delta x = \ell_0 / 512 \approx 8 \times 10^3$ AU is significantly larger than the dissipation scale provided by the kinematic viscosity of interstellar gas of $\ell_d \sim 10$ AU \citep{jou1998}. If the turbulence were allowed to cascade down to these small scales, the distribution of the heating rate would be more intermittent than it is in our simulation \citep{pan2009}. However, viscous dissipation is not the only process that removes energy from the turbulent cascade. Ion-neutral friction becomes significant at the much larger scale $\ell_{\rm{AD}}$ (see Section 2.4), and is capable of dissipating most ($\approx 70$ \% for $\mathcal{M}_A \approx 1$) of the energy in the cascade at scales $\approx$  $10$  $\ell_{\rm{AD}}$ \citep{li2012}, which is comparable to the cell size in our numerical data. Below $\ell_{\rm{AD}}$, a turbulent cascade can re-assert itself in the neutrals, allowing some of the energy to be dissipated on the $\sim$ 10 AU scale set by viscosity. While our procedure likely under-estimates the intermittency for this remaining $\approx 30$ \% of the energy, it is more accurate than the opposite assumption that all the energy in cascade makes it to $\sim 10$ AU scales. Note that this procedure for estimating the turbulent heating rate is scaled such that it includes $\emph{all}$ the energy removed from the turbulent cascade by dissipative processes, whether the physical mechanism is molecular viscosity or ambipolar diffusion. We do not separately include an ambipolar diffusion heating term in Equation (\ref{eq:ebalance}) because doing so would amount to double counting.

For the C$^+$ and O cooling coefficients, we adopt the formulas given by \citet{wol2003}:
\begin{align}
\lambda_{\rm C\hspace{-1pt}^+}(T) &= 3.6 \times 10^{-27}  \exp{(-92  \hspace{3 pt} {\rm{K}} / T)} \hspace{3 pt} \rm{erg} \hspace{3 pt} \rm{cm}^3 \hspace{3 pt} \rm{s}^{-1} \nonumber \\
\lambda_{\rm O}(T) &= 2.35 \times 10^{-27} \left( \frac{T}{100 \hspace{3 pt} {\rm{K}}} \right)^{0.4} \nonumber \\
&\times \exp{(-228  \hspace{3 pt} {\rm{K}} / T)} \hspace{3 pt} \rm{erg} \hspace{3 pt} \rm{cm}^3 \hspace{3 pt} \rm{s}^{-1}, 
\end{align} 
where we have scaled the overall numerical factors to account for our fractional abundances of carbon and oxygen of $1.6 \times 10^{-4}$ and $3.2 \times 10^{-4}$, rather than the $1.4 \times 10^{-4}$ and $3.4 \times 10^{-4}$ used in \cite{wol2003}. The cooling rates per H nucleus are then $\Lambda_{\rm C^+}(T) = n_{\rm{H}} \lambda_{\rm C^+}(T)$ and $\Lambda_{\rm O}(T) = n_{\rm{H}} \lambda_{\rm O}(T)$. These rates are valid for $n_{\rm{H}} < n_{\rm{crit}} \simeq 3000$ cm$^{-3}$.

The final thermal process we consider is the cooling rate due to the H$_2$ molecule. This coolant is particularly important in the warm (over a few hundred K) gas in which CH$^{+}$ is expected to form. In the low density limit, the H$_2$ level populations are sub-thermal and the cooling rate $\Lambda_{\rm{H}_2}(n_{\rm H} \rightarrow 0)(T)$ is a sum over the rates due to collisions with H, H$_2$, He, and e. For these rates, we use the tables in \cite{glover2008}, assuming an ortho:para ratio of 0.7 (see section 3.3) and the fractional abundances of H, H$_2$, He and electrons listed in Table \ref{tbl-1}. These rates are valid at arbitrarily low gas densities, since the cooling rate coefficients themselves are independent of the gas density in this limit. At densities high enough for local thermodynamic equilibrium to be established, the H$_2$ cooling rate per H nucleus $\Lambda_{\rm{H}_2 , \hspace{3 pt} \rm{ LTE}}(T)$ becomes independent of the collision partner abundances, and we adopt the cooling rate from \cite{coppola2012}. We bridge these two limits following \cite{hol1979}:
\beq
\Lambda_{\rm{H}_2}(T) = \frac{\Lambda_{\rm{H}_2 , \hspace{3 pt} \rm{ LTE}}(T)}{1 + \Lambda_{\rm{H}_2 , \hspace{3 pt} \rm{ LTE}}(T) / \Lambda_{\rm{H}_2}(n_{\rm H} \rightarrow 0)(T)}
\eeq
The total cooling rate is then 
\beq
\Lambda_{\rm{tot}}(T) = \Lambda_{\rm C\hspace{-1pt}^+}(T) + \Lambda_{\rm O}(T) + x_{\rm{H}_2} \Lambda_{\rm{H}_2}(T).
\eeq Equation (\ref{eq:ebalance}) becomes a non-linear equation for $T$ in each cell, which we solve numerically using the {\tt scipy.optimize.brenth} routine from the SciPy software library \citep{SciPy}. The magnitudes of these cooling processes are summarized as a function of temperature in Figure \ref{fig:energy_balance} for our standard model parameters. For comparison, the (constant) values of the various heating rates are also displayed for our standard density of $n_{\rm H} = 30$ cm$^{-3}$. 

We estimate the typical cooling time in our model as the thermal energy density divided by the cooling rate:
\begin{equation}
\label{eq:cooling_time}
t_{\rm{cool}} = \frac{3/2 k T}{\Lambda_{\rm{tot}}(T) } \simeq 3.3 \times 10^{4} \rm{yr},
\end{equation} where the numerical evaluation is for our fiducial values of $n_{\rm{H}} = 30$ cm$^{-3}$ and $T_{50, \rm{M}} = 35$ K. This is two orders of magnitude smaller than the dynamical time (Equation (\ref{eq:dynamical_time})), which justifies our use of an energy balance equation.

\subsection{Ion-Neutral Drift}

\begin{figure}
  \centering
    \includegraphics[width=0.45\textwidth]{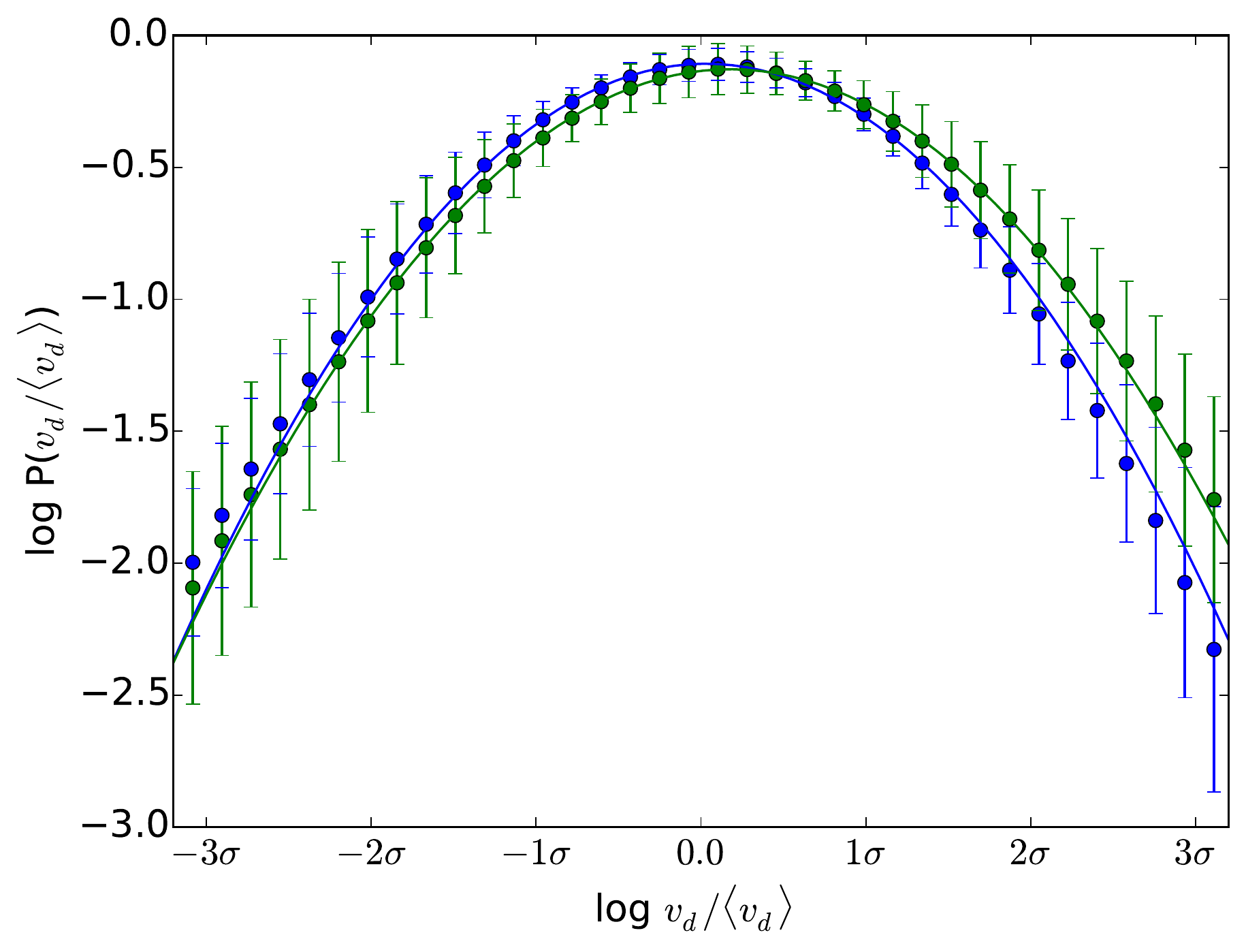}
    \caption{\label{fig:drift_compare} Blue - the circles show the mass-weighted distribution of $v_d$ divided by its mean value $\langle v_d \rangle$ from the $\mathcal{M} = 3$, $\mathcal{M}_{\rm{A}} = 0.67$, $\radlo \approx 1000$ AD simulation. The error bars show the $2 \sigma$ temporal variation in distribution over 2 box crossing times, and solid line shows the best-fit log-normal. Green - same, but for the $\mathcal{M} = 3$, $\mathcal{M}_{\rm{A}} = 0.67$ ideal simulation, with $v_d$ computed from Equation (\ref{eq:lorentz_drift}). The agreement between the two curves is quite good over more than 3 standard deviations.}
\end{figure}

\begin{figure}
  \centering
    \includegraphics[width=0.45\textwidth]{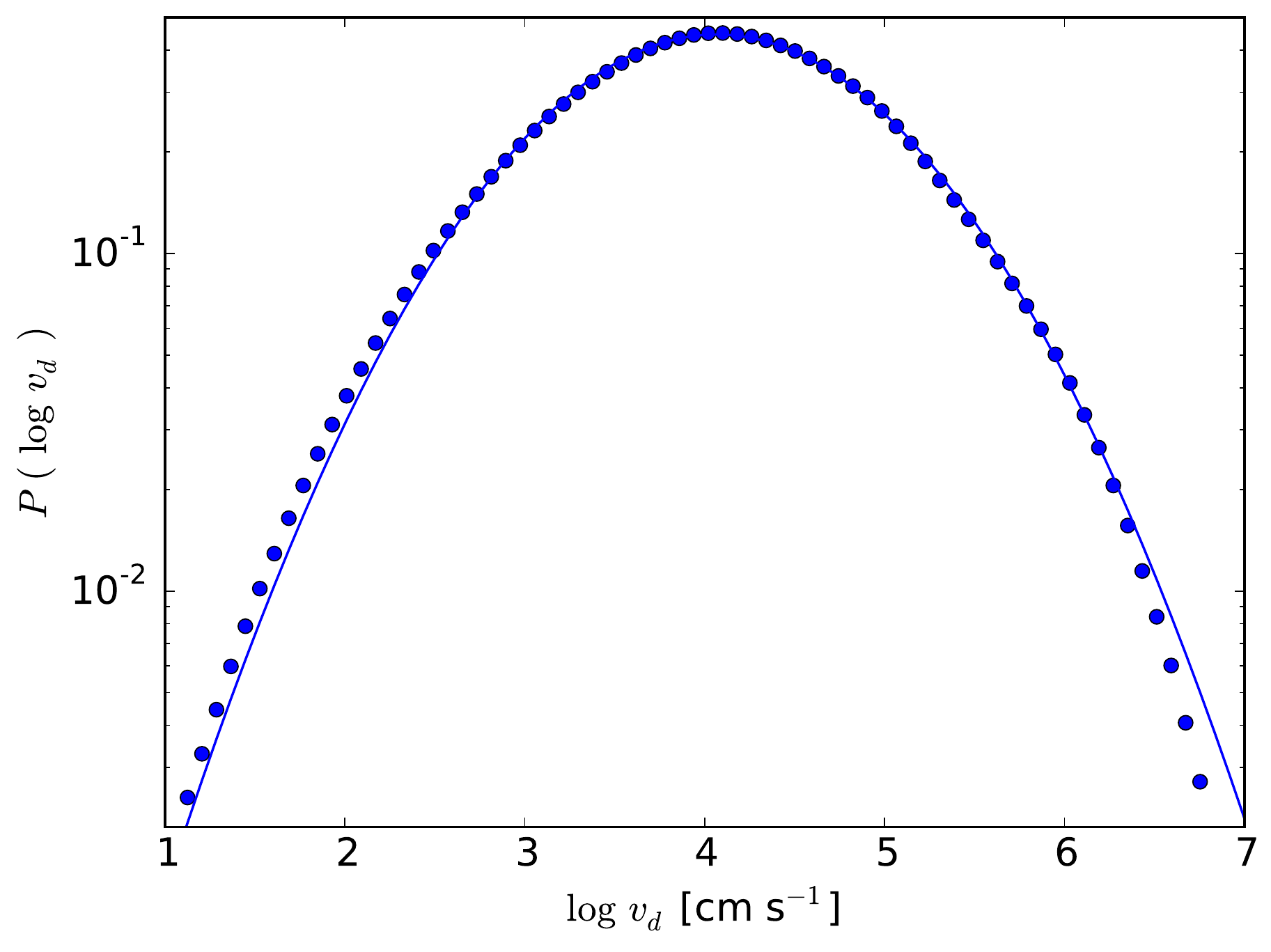}
    \caption{\label{fig:drift} Distribution of $\log v_d$ for our model in physical units. The blue circles are the simulation data, and the solid line is a best-fit normal distribution with a mean of $4.04$ and a standard deviation of $0.89$.}
\end{figure}

In the presence of ambipolar diffusion (the net slippage between the charged and neutral species in a plasma), ion-neutral reactions like (1) can proceed at rates faster than those expected from the kinetic temperature alone \citep[e.g.][]{dra1980, flower1985}. The relative importance of the ambipolar and inertial forces in a turbulent system can be characterized by the ambipolar diffusion Reynolds number $\radlo$ \citep{zwe1997, li2012}:
\beq
\radlo \equiv \frac{4 \pi \gad \rho_i \rho_n \ell_0 \sigma_{\rm 3D}} {B_{\rm rms}^2},
\label{eq:radell} 
\eeq  where $\rho_i$ and $\rho_n$ are the densities of the ionic and neutral components of the fluid and $\gad$ is the ion-neutral coupling constant given by $\langle \sigma v \rangle / (m_i + m_n)$. For C$^+$ and H$_2$, this evaluates to $8.47 \times10^{13}$ cm$^3$ s$^{-1}$ g$^{-1}$ \citep{dra1980}. Applying Equations (\ref{eq:ls}) and (\ref{eq:column}) for our adopted degree of ionization and magnetic field strength, $\radlo$ is
\beq
\label{eq:radnumber}
\radlo \approx 6.3 \times 10^3 \sqrt{\left(  \frac{n_{\rm{H}}}{30 \hspace{3 pt} \rm{cm}^{-3}} \right) } \gg 1.
\eeq The corresponding length scale $\ell_{\rm{AD}} = \ell_0 / \radlo$ at which ambipolar dissipation becomes significant is $\approx 640$ AU for $n_{\rm{H}} = 30$ cm$^{-3}$. Thus, ambipolar drift should not be significant on large scales in our model. However, as with the turbulent heating rate, there may be isolated regions in the tails of the drift velocity distribution where this effect is significant. 

To proceed, we need a prescription for computing $v_d$. Unfortunately, two-fluid simulations of MHD turbulence are prohibitively expensive in the high $\mathcal{M}$, strongly coupled regime considered here. To estimate the effects of $v_d$ on the production of CH$^+$, we instead use our ideal MHD data along with an approximate analytic expression for the drift velocity in the strongly coupled regime, which we corroborate with direct numerical simulations of turbulent ambipolar diffusion at lower $\mathcal{M}$. Specifically, if the system is weakly ionized, then the Lorentz force and the ion-neutral drag force dominate all the other terms in the ion momentum equation and the drift is given by $v_d = | ( \vec \nabla \times \vec B ) \times \vec B | / 4 \pi \gamma_{\rm{AD}} \rho_i \rho_n$ \cite[e.g.][]{shu1992}. If the effects of ambipolar diffusion are weak enough that they have only a minor effect on the geometry of the magnetic field, we can estimate the drift by computing $| ( \vec \nabla \times \vec B ) \times \vec B |$ in the ideal limit:
\beq
\label{eq:lorentz_drift}
v_d \approx \frac{ | ( \vec \nabla \times \vec B ) \times \vec B |^{\rm{Ideal}} }{4 \pi \gamma_{\rm{AD}} \rho_i \rho_n}.
\eeq

This procedure is illustrated in Figure \ref{fig:drift_compare}. We take two simulations of $\mathcal{M} = 3$, $\mathcal{M}_{\rm{A}} = 0.67$ turbulence from \cite{li2008}, one which follows the ion and neutral fluids separately, and one which assumes ideal MHD. The AD simulation has $R_{\rm{AD}}(\ell_0) \approx 1000$. We directly compute the time-averaged, density-weighted distribution of $v_d$ in the non-ideal simulation, and compare it to that of Equation (\ref{eq:lorentz_drift}) computed using the ideal data with $\gamma_{\rm{AD}}$ and $\chi_i$ chosen to match the AD simulation. The resulting distributions both have an approximately log-normal form:
\begin{align}
\label{eq:normal}
P(\log v_d) &d \log v_d = \nonumber \\
&\frac{1}{\sigma_{\log v_d} \sqrt{2 \pi}} \exp \left( - \frac{ (\log v_d - \mu_{\log v_d})^2 }{2 \sigma_{\log v_d}^2} \right)
\end{align}
and agree with each other to within the error bars, which show the magnitude of the temporal fluctuation in the drift distribution computed over 2 crossing times. Equation (\ref{eq:lorentz_drift}) does slightly over-predict the simulated value of $v_d$: both the mean $\mu_{\log v_d}$ and standard deviation $\sigma_{\log v_d}$ of the Lorentz drift approximation are larger than those of
the log-normal fit to the true drift distribution by 5\% and 2\%, respectively. This is likely because, although $R_{\rm{AD}}(\ell_0)$ for the entire box is $\sim 1000$, there are still sub-regions in the box where the coupling less strong. We expect that this approximation should improve with increasing $R_{\rm{AD}}(\ell_0)$. 

Figure \ref{fig:drift} shows the result of applying this procedure to our $\mathcal{M} \approx 10$, $\mathcal{M}_{\rm{A}} \approx 2.2$ Ideal MHD data, scaled to physical units as described above. Here, we again find that the distribution of $\log (v_d)$ is approximately normal, with best-fit parameters $\mu_{\log v_d} = 4.04$ and  $\sigma_{\log v_d} = 0.89$. Although the error in the best-fit lognormal parameters found above is small, it could potentially have a large impact on the CH$^+$ abundance, since that is primarily determined by the tails of the distribution. We can estimate the accuracy of our approximation as follows. First, we compute the mean CH$^+$ abundance (see Section 2.5 below) in a cloud with constant density $n_{\rm H} = 30$ cm$^{-3}$ and kinetic temperature $T = 35$ K using Equations (\ref{eq:abundance}) and (\ref{eq:Teff}), under the assumption that the distribution of $v_d$ is given by Equation (\ref{eq:normal}) with our best fit parameters. We then recompute the CH$^+$ abundance using values of $\mu_{\log v_d} $ and  $\sigma_{\log v_d} $ that are lower by 5\% and 2\%, respectively - the error we found for the $\mathcal{M} = 3$, $\mathcal{M}_{\rm{A}} = 0.67$ case above. The result is that the two abundances agree to within a factor of 2.  As the $R_{\rm{AD}}(\ell_0)$ in our target system is larger than 1000 (Equation \ref{eq:radnumber}), we expect the true error to be somewhat less than this. 

\subsection{Chemistry} 
\label{sec:chemistry}
To assess the viability of turbulent dissipation as an energy source for CH$^+$ production, we perform a simple analytic estimate of the CH$^+$ abundance in a cell as a function of $n_{\rm H}$ and $T_{\rm{eff}}$ following \cite{lambert86}. CH$^+$ forms through reaction \ref{reaction} with a rate constant: 
\beq
\label{rate}
k_{\rm f} = 1.5 \times 10^{-10} \times \exp (-4640 \hspace{ 3 pt} \mbox{K} / T_{\rm{eff}} ).
\eeq 
Once it forms, it will be quickly destroyed by reactions with H, H$_2$, and electrons. The rates for these process are 
\beqa
\label{reaction_network}
	\rm{CH}^+ + \rm{H} \rightarrow \rm{C}^+ + \rm{H}_2 , 
	 & k_{\rm{H I}}= 1.5 \times 10^{-10} \mbox{ cm$^3$ s\e} \nonumber \\ 
	 \rm{CH}^+ + \rm{H}_2 \rightarrow \rm{CH}_2^+ + \rm{H},
	 & k_{\rm{H}_2} = 1.2 \times 10^{-9} \mbox{ cm$^3$ s\e} \nonumber \\ 
	 \rm{CH}^+ + \rm{e} \rightarrow \rm{C} + \rm{H},
	 & k_{\rm{e}} = 1.5 \times 10^{-7}\mbox{ cm$^3$ s\e},
\eeqa 
where we have adopted the values used by the Meudon PDR code\footnote{http://pdr.obspm.fr/PDRcode.html}.

Because the electron fraction $x_e \approx x_i$ is $\sim 10^{-4}$, removal of CH$^+$ by electrons is not crucial and we ignore it in our calculations. However, destruction by atomic and molecular hydrogen are both important. Balancing the rate of formation with the rate of destruction $n_{\rm{C}\hspace{-1pt}^+} n_{\rm{H}_2} k_f = n_{\rm{CH}\hspace{-1pt}^+} n_{\rm{H I}} k_{\rm{H I}} + n_{\rm{CH}\hspace{-1pt}^+} n_{\rm{H}_2} k_{\rm{H}_2} $, we can derive 
\beqa
\label{eq:abundance}
n_{\rm{CH}\hspace{-1pt}^+} &= x({\rm{C}}^+) \frac{ x(\rm{H}_2) }{1 - 2 x(\rm{H}_2)} \frac{k_{\rm f}}{k_{\rm{HI}}} \left( 1 + \frac{k_{\rm{H}_2}}{k_{\rm{HI}}} \frac{x(\rm{H}_2)}{1 - 2 x(\rm{H}_2)} \right)^{-1} n_{\rm{H}} \nonumber \\
&\approx  3.9 \times 10^{-4} \left( \frac{n_{\rm{H}}}{30 \hspace{3 pt} \rm{cm}^{-3}} \right) \times \rm{exp} \left( \frac{-4640 \hspace{ 3 pt} \rm{K}}{\it{T}_{\rm{eff}}} \right).
\eeqa  Equation \ref{eq:abundance} highlights the importance of both the molecular fraction and the C$^+$ abundance for CH$^+$ production; the gas must be well-shielded enough from the ambient radiation field that some of the hydrogen is molecular form, but not so well-shielded that carbon is all molecular.

We can also estimate a typical CH$^+$ formation timescale as follows. Suppose we are in a region that initially contains no CH$^+$ that is subsequently subject to heating. From reaction \ref{reaction_network}, the rate of change in $n_{\rm{CH}\hspace{-1pt}^+}$ is 
\beq
\frac{d n_{\rm{CH}\hspace{-1pt}^+}}{dt} = n_{\rm{C}\hspace{-1pt}^+} n_{\rm{H}_2} k_f - n_{\rm{CH}\hspace{-1pt}^+} n_{\rm{H I}} k_{\rm{H I}} - n_{\rm{CH}\hspace{-1pt}^+} n_{\rm{H}_2} k_{\rm{H}_2}.
\eeq The solution to this equation is
\beq
n_{\rm{CH}\hspace{-1pt}^+}(t) = n_{\rm{CH}\hspace{-1pt}^+, \rm{eq}} \times \left[ 1 - \exp \left( {-a t} \right) \right],
\eeq where $n_{\rm{CH}\hspace{-1pt}^+, \rm{eq}}$ is the equilibrium abundance given by Equation \ref{eq:abundance} and $a = (1 - 2 x_{\rm{H}_2}) k_{\rm{H I}} + x_{\rm{H}_2} k_{\rm{H}_2}$. Thus, the time over which the CH$^+$ abundance achieves 90\% of its equilibrium value is
\beq
t_{\rm{CH}^+} = \frac{\ln(10)}{a} \approx 250 \hspace{3 pt} \rm{yr}. 
\eeq This is 2 orders of magnitude shorter than the cooling time (Equation \ref{eq:cooling_time}), and 4 orders shorter than the characteristic dynamical time (Equation \ref{eq:dynamical_time}).

\section{Results}
\label{sec:results}

\subsection{Temperature and CH$^+$ Abundance}

\begin{figure}
  \centering
    \includegraphics[width=0.45\textwidth]{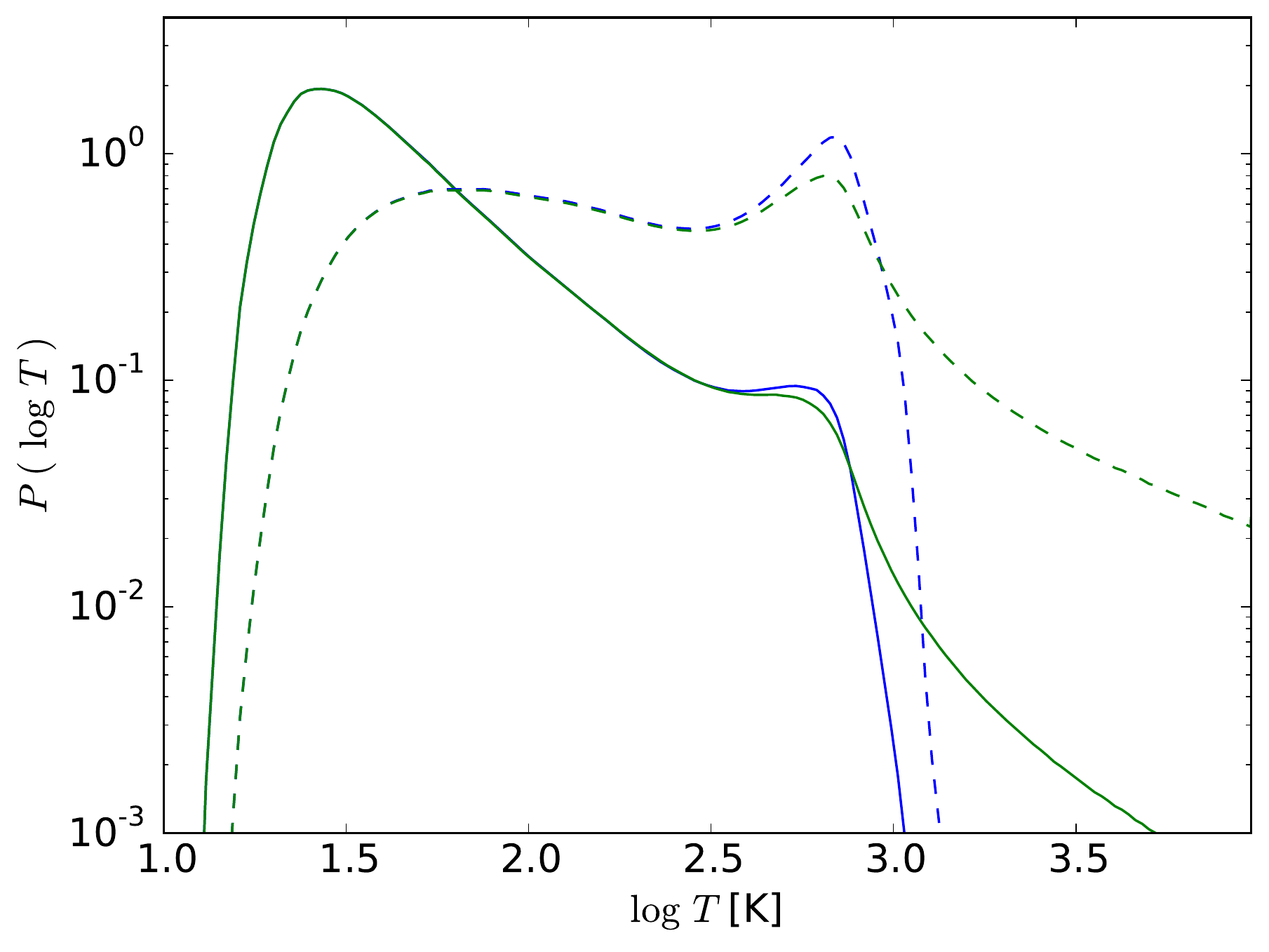}
    \caption{\label{fig:PDF} Blue solid line - mass-weighted differential distribution function of $\log T$. Green solid line - the volume-weighted distribution of same. Dotted lines - the colors have the same meaning as before, but the distribution of $\log T_{\rm{eff}}$ has been plotted instead.}
\end{figure}

\begin{figure}
  \centering
    \includegraphics[width=0.45\textwidth]{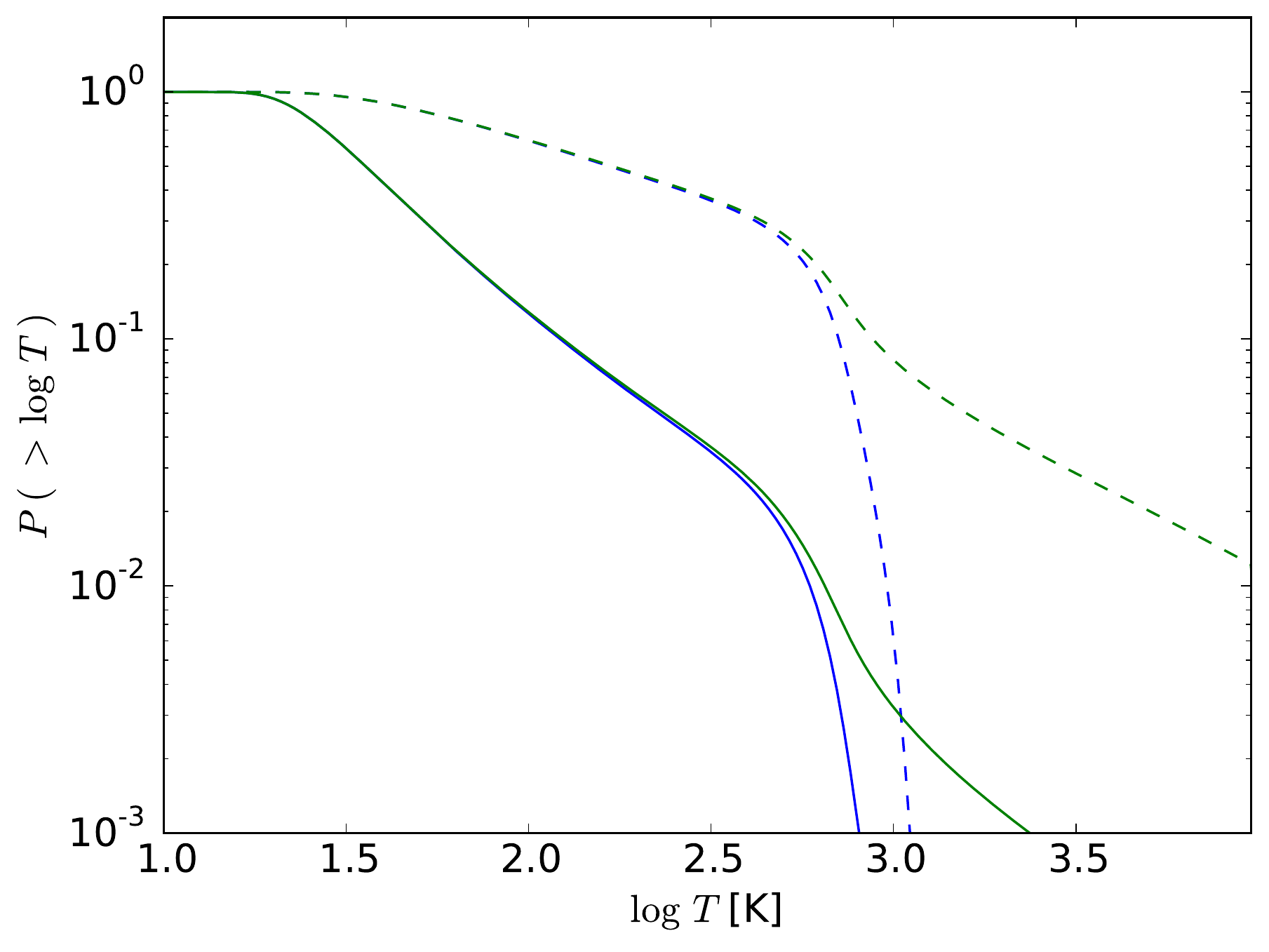}
    \caption{\label{fig:CDF} Same as Figure \ref{fig:PDF}, but showing the cumulative rather than the differential distribution functions. Specifically, $P$ $($ $>$ $\log$ $T$ $)$ shows the fraction of the mass or volume with values of $\log T$ greater than the corresponding value on the x-axis. As before, the solid lines show the distribution of $\log T$ and the dotted lines the distribution of  $\log T_{\rm{eff}}$. }
\end{figure}

The result of our temperature calculation is summarized in Figures \ref{fig:PDF} and \ref{fig:CDF}. Figure \ref{fig:PDF} shows the mass- and volume- weighted differential probability distribution functions of $\log T$ and $\log T_{\rm{eff}}$. Figure \ref{fig:CDF} shows the cumulative distribution functions of the same quantities. Both $T$ and $T_{\rm{eff}}$ can take a large range of values spanning over 3 orders of magnitude. The distributions of both quantities appear to be bimodal, showing peaks at several tens and and several hundreds of K. The large majority of the mass lies at low $T$, but a small fraction is found in high temperature ``pockets" - precisely the arrangement proposed by \cite{falgaronepuget1995}. An interesting feature of these distributions is the importance of density variations in setting the temperature. The difference between the mass- and volume weighted distributions shows that high values of $T$ and $T_{\rm{eff}}$ tend to be found in relatively low-density gas. This is easily understood from Equations (\ref{eq:ebalance}) and (\ref{eq:lorentz_drift}). From Equation (\ref{eq:ebalance}), the heating processes are all proportional to $n_{\rm{H}}$. Furthermore, at low $T$, the dominant cooling processes are C$^+$ and O lines, for which the cooling power goes like $n_{\rm{H}}^2$. Accordingly, cooling can balance heating at relatively low temperatures unless $n_{\rm{H}}$ is small. Similarly, the drift velocity $v_d$ is inversely proportional to $n_{\rm H}^2$ for a fixed ionization fraction (Equation \ref{eq:lorentz_drift}), meaning that it is likely to be small except in low density regions. In our calculations, we find that the volume-weighted mean temperature $\bar T_{\rm V} \approx 290$ K exceeds the density-weighted value $\bar T_{\rm M} \sim 67$ K by more than a factor of 4. Similarly, the volume-weighted median $T_{50, \rm{V}} \approx 163$ K exceeds the mass-weighted median $T_{50, \rm{M}} \approx 35$ K by a similar value. This difference is particularly dramatic for the high-temperature tails of the distribution: while approximately 8\% of the volume in our box has a $T_{\rm{eff}}$ greater than 1000 K, only 0.3\% of the mass does. 

\begin{figure*}
  \centering
    \includegraphics[width=1.0\textwidth]{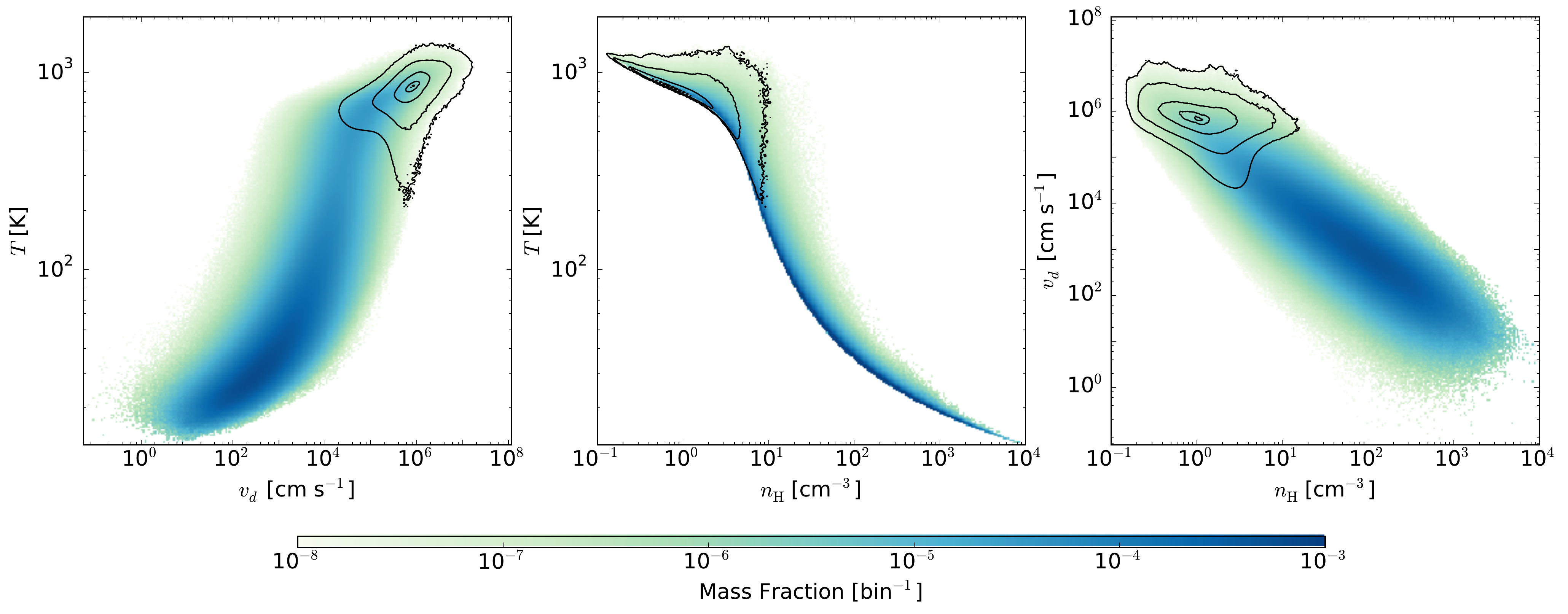}
    \caption{ \label{fig:phase} Left - the color scale shows the fraction of the total mass in each logarithmically-spaced $T - v_d$ bin. The black contours show the region of phase space in which 99, 90, 50, 10, and 1 percent of the CH$^+$ is found. Middle - same, but for $T - n_{\rm H}$. Right - same, but for $v_d - n_{\rm H}$. }
\end{figure*}

We thus expect CH$^+$ to mostly be found in low-density regions. We show in Figure \ref{fig:phase} the regions of $T - v_d$, $T - n_{\rm H}$, and $v_d - n_{\rm H}$ phase space in which most of the CH$^+$ mass is contained. For instance, in the middle panel, we have created 256$^2$ logarithmically-spaced bins spanning the full range of temperatures and densities found in the simulation output. The color scale shows the fraction of the total mass in each of these $T - n_{\rm H}$ bins, while the black contours show the regions of phase space that contain the top 99\%, 90\%, 50\%, 10\%, and 1\% of the the CH$^+$. The left and right panels repeat this procedure for $T - v_d$ and $v_d - n_{\rm H}$, respectively. These plots confirm our expectation that regions with $T \gtrsim 1000$ K and $v_d \gtrsim \sigma_{3 \rm{D}} \approx 4$ km s$^{-1}$ tend to be found at low density, with typical values of $n_{\rm H} \sim$ a few H nuclei per cm$^3$. They also show that, while these regions are rare, they contain almost all of the CH$^+$ molecules. 

The middle panel of Figure \ref{fig:phase} illustrates the importance of the intermittency in $\Gamma_{\rm{Turb}}$ in setting the gas temperature. The lower curve of this diagram traces out a line that corresponds to what the temperature as a function of density would be if we considered only $\Gamma_{\rm{PE}}$, $\Gamma_{\rm{CR}}$, and the various cooling processes described in section 2.3. Most of the gas in the simulation receives little heating from turbulence dissipation and thus lies at or near this line. A small fraction of the gas, however, is heated to significantly higher temperatures by the effects of turbulence, and this gas comprises the bulk of the material that is heated above $\sim 1000$ K. We find that without $\Gamma_{\rm{Turb}}$, the fraction of mass in the simulation heated to $T > 1000$ K would drop by approximately a factor of 20.  

\cite{she2008}, following \cite{rit2006}, used the ratio of $N_{\rm{CH}^+}$ to $N_{\rm{CH}}$ along with $G_0$ and $x(\rm{H}_2)$ as an empirical probe of the gas density along a number of diffuse sight lines. The resulting density estimates were generally quite low, with typical $n_{\rm H} \sim 3$ cm$^{-3}$, and in some cases much lower than estimates for $n_{\rm H}$ inferred from C$^+$ excitation for the same sight lines \citep{son2002, son2003}. \cite{she2008} interpret this as saying that a significant portion of the extinction and atomic hydrogen are associated with purely atomic regions that contain no CH$^+$, and that the corresponding increase in $G_0$ increases the estimate for $n_{\rm H}$. Our result suggests an alternative explanation: that the CH$^+$ really is predominately found at low $n_{\rm H}$, while the C$^+$ observations are probing something closer to the mean density.

\subsection{Sight-line Analysis}

\begin{figure}
  \centering
    \includegraphics[width=0.45\textwidth]{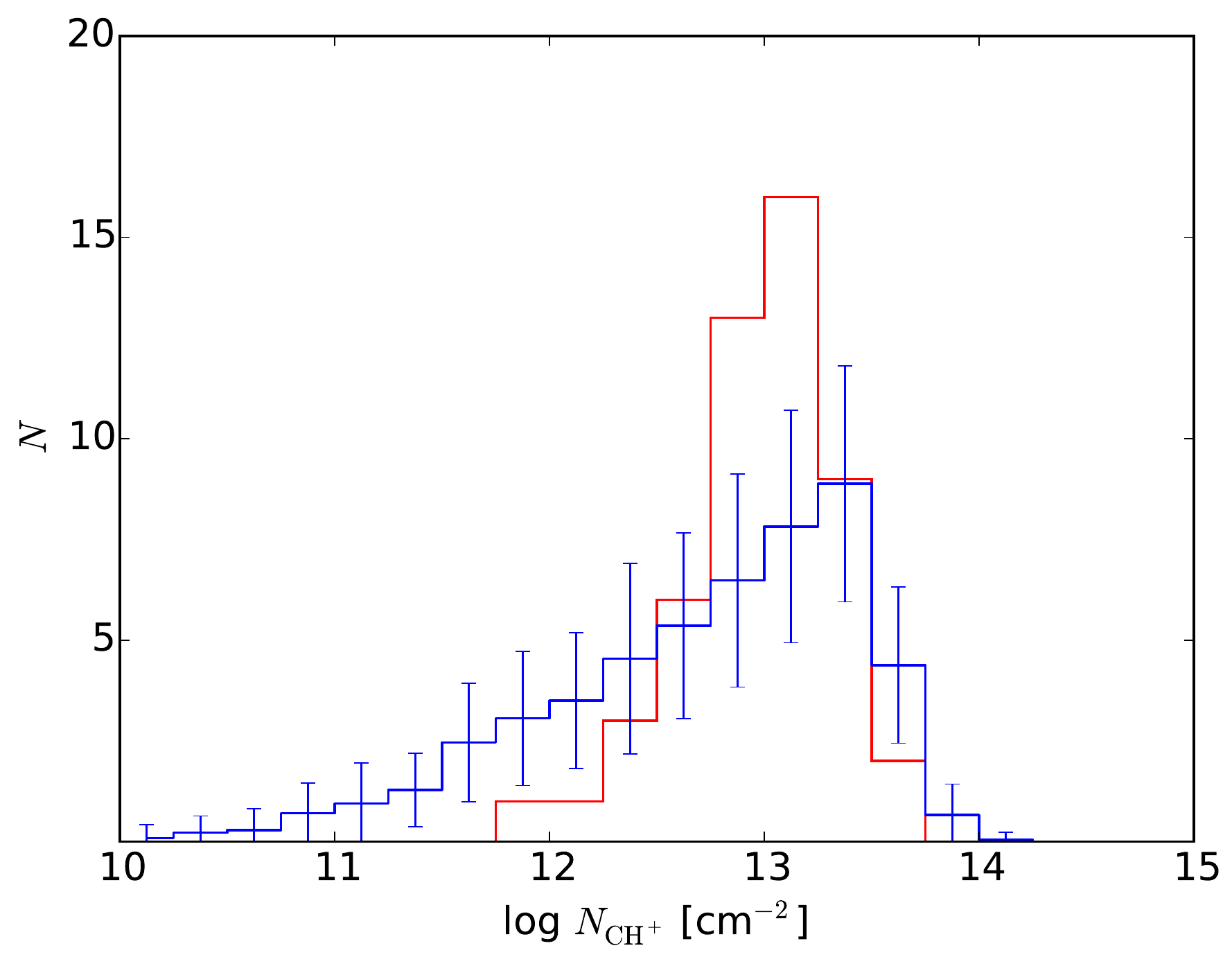}
    \caption{\label{fig:all_rays} Blue - average histogram of CH$^+$ column densities over 2500 lines of sight passing through the simulation volume. The error bars indicate the $1-\sigma$ range of variation over all 50 samples of 50 sight lines each. Red - histogram of the CH$^+$ column densities from the 50 sight line sample in \protect\cite{wes2008}.}
\end{figure}

\begin{figure}
  \centering
    \includegraphics[width=0.45\textwidth]{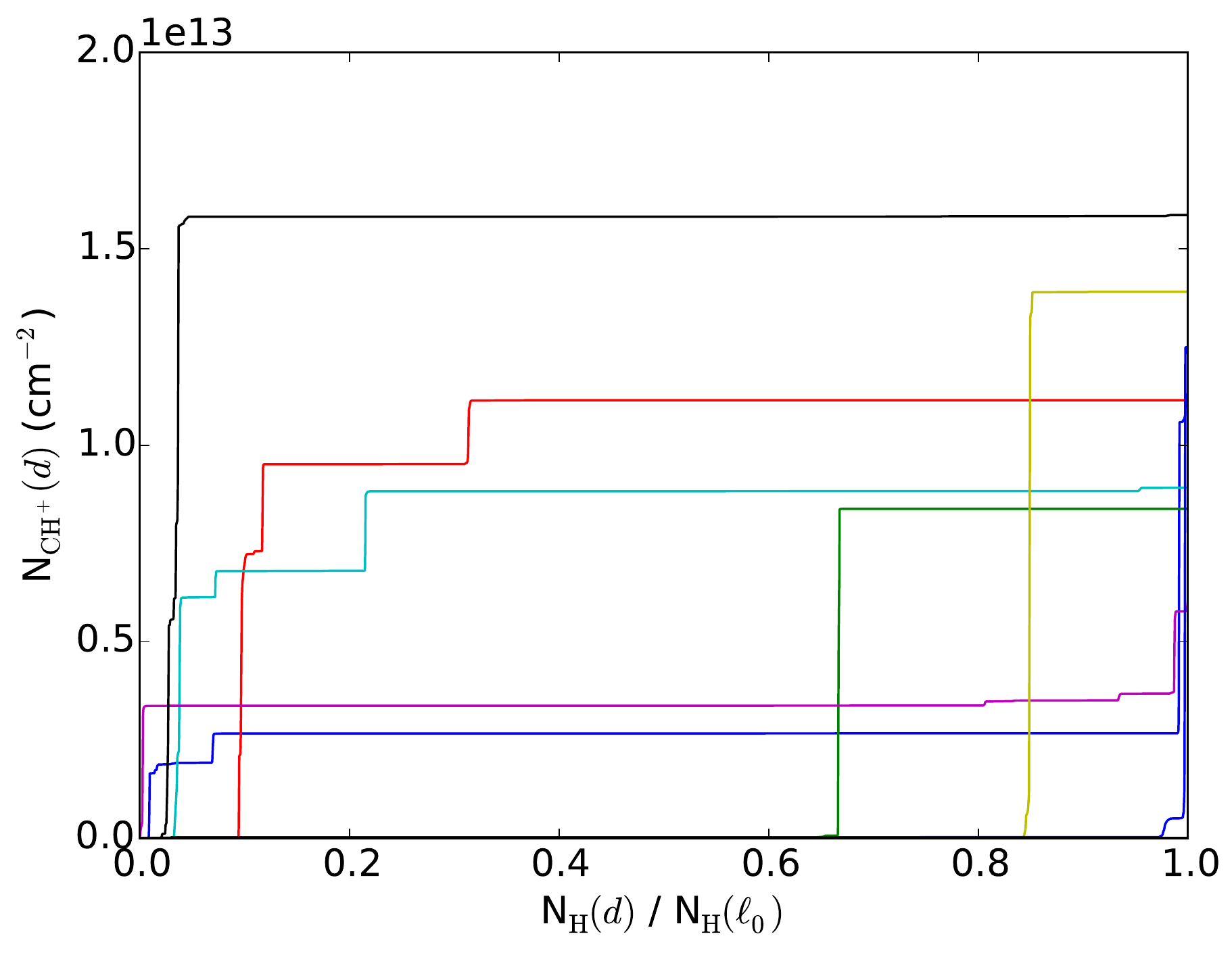}
    \caption{\label{fig:eight_rays} CH$^+$ column density versus total column density $N_{\rm{H}}$ integrated along the same eight sight lines shown in Table \ref{tbl-2}. The quantities $N_{\rm{CH}^+}(d)$ and $N_{\rm H }(d)$ are the CH$^+$ and total column densities integrated up though path length $d$ along each ray. The $x$-axis has been normalized by the total column density $N_{\rm H}(\ell_0)$ to fit all the rays on the same plot.}
\end{figure}

\cite{wes2008} presented a sample of 53 CH$^+$-containing sight lines, 50 of which also had measurements of $N_{\rm H}$ and $N_{\rm{H}_2}$.  For the purpose of comparing with observations, we will adopt these 50 sight lines as our observational sample. We limit ourselves to this sample so that the CH$^+$ column densities would all be calculated in a consistent way, but the distribution of columns in \cite{wes2008} is quite similar to that of \cite{she2008}, which included both original data and results from many previous studies \citep{fed1997, gre1997, knauth2001, pan2004}. 

Our model, by construction, has the same mean values of $\bar N_{\rm H}$ and $\bar N_{\rm{H}_2}$ as the \cite{wes2008} sample. The observed mean and median CH$^+$ column densities for these sight lines are $1.2 \times 10^{13}$ cm$^{-2}$ and $1.1 \times 10^{13}$ cm$^{-2}$, respectively. The corresponding values in our model are $1.3 \times 10^{13}$ cm$^{-2}$ and $8.1 \times 10^{12}$ cm$^{-2}$, i.e. our mean is higher by 8\% and our median is lower by 26\%.  To further compare against the observational sample, we generate 2500 synthetic observations by casting 50 groups of 50 rays orthogonally through our computational domain as follows. For each ray, we randomly select one of the $x$, $y$, and $z$ directions, and then we randomly select a coordinate describing ray's position in the corresponding normal plane. For example, for a ray aligned with the $z$-axis, we would randomly select a point in the $xy$ plane from a uniform distribution to describe the position of the ray in the box. For each group of rays, we construct a histogram of $N_{\rm{CH}^+}$ using 20 bins spaced evenly of the range $\log N_{\rm{CH}^+} = 10$ to $\log{N_{\rm{CH}^+} = 15}$ and compare it against the corresponding histogram for the observational sample. The result is shown in Figure \ref{fig:all_rays}. The error bars indicate the $1-\sigma$ variation in the number of sight lines per bin. The number of groups was set at 50 because that number was sufficient for the error bars to be converged: the maximum change in the $1-\sigma$ error over all the bins was 15\%, and the mean change was only 1\%. We find that, while our model agrees well with the mean and median of the observational sample, it does tend to over-produce both very large and very small values. However, the sight lines in \cite{wes2008} do not constitute a truly random sampling in that they were chosen because CH$^+$ lines were detectable and the H and H$_2$ column densities were measurable. This is likely not the case for the very low column lines in our model. 

From our sample of 2500 rays, we choose 8 ``typical" sight lines for more detailed investigation as follows. First, we randomly draw a ray from the sample. Next, we keep it only if both $N_{\rm{CH}^+}$ and $N_{\rm H}$ are within 50\% of the overall sample means. Otherwise, we throw it away and draw again. We stop when 8 rays have been selected. The properties of these sight lines are described in Table \ref{tbl-2}. Here, the notation $\bar x_{99}$ means the average value of $x$ in the cells that are in the top 99\% of the CH$^+$ number density distribution. Thus, $\bar n_{\rm{H},99}$, $\bar T_{\rm{M}, 99}$, and $\bar v_{d, 99}$ are the mean density, mass-weighted temperature, and drift velocity in the regions which contain 99\% of the CH$^+$. We find that, consistent with Figure \ref{fig:phase}, almost all of the CH$^+$ is in these sight lines is found in low-density, high-temperature, high-drift pockets of gas, with typical values of $\bar n_{\rm{H},99} \approx 1 - 2$ cm$^{-3}$, $\bar T_{\rm{M}, 99} \approx 700$ K, and $\bar v_{d, 99} \approx 2 - 3$ km s$^{-1}$. These temperatures and velocities are quite similar to those obtained in the TDR models of \cite{god2009} using very different techniques.   

The ``temperature" $\mu_{\rm H} v_d^2 / 3 k$ associated with a 4 km s$^{-1}$ drift velocity is $\approx 900$ K, comparable to $\bar T_{\rm{M}, 99}$. Both effects thus appear to be important for building up CH$^+$ columns in excess of $10^{13}$ cm$^{-2}$. To gauge the relative importance of the two effects, we repeat our calculation with $v_d$ computed as above, but with $T$ fixed at 35 K. The result is that the mean CH$^+$ column drops from $\approx 1.3 \times 10^{13}$ cm$^{-2}$ to $\approx 8.0 \times 10^{12}$ cm$^{-2}$. If we repeat the same experiment with $T$ computed as above, but ignoring the effects of $v_d$, the CH$^+$ column drops dramatically, down to $\approx 5.0 \times 10^{11}$ cm$^{-2}$. Thus, the CH$^+$ chemistry in our model appears to be mainly driven by ion-neutral drift, with the kinetic temperature making a secondary, but not negligible, contribution to the total column. 

Finally, to gauge the spatial extent of these regions, we show in Figure \ref{fig:eight_rays} the result of integrating $N_{\rm{CH}^+}$ and $N_{\rm H}$ along each of the 8 rays in Table \ref{tbl-2}. All of the rays show the same general behavior: there are large regions that make basically no contribution to CH$^+$ column, punctuated by a few thin zones where the CH$^+$ abundance is substantial. The typical sight line intersects approximately 2-4 of these regions. This analysis further confirms the view of \cite{falgaronepuget1995} - that the cold ISM contains isolated patches of hot, chemically active gas, and that these regions are crucial to understanding diffuse cloud chemistry.

The gas densities we infer for the CH$^+$-containing regions in our model are lower than those typically associated with gas containing substantial abundances of H$_2$. As discussed in Section \ref{sec:chemistry}, a substantial H$_2$ fraction is crucial for CH$^+$ formation, and our model suggests that low densities are crucial, as well. However, it is not only the local gas density that is important for setting the H$_2$ fraction; the total shielding from the ambient radiation field matters as well. In PDR models that assume constant density, this distinction is not made, but in a supersonically turbulent medium, it is entirely possible for a region with a low local gas density to nonetheless be well-shielded from FUV radiation. Indeed, Table \ref{tbl-2} and Figure \ref{fig:eight_rays} show that, in our model, the CH$^+$-forming regions tend to be randomly distributed along sight lines, so that many of them would be well-shielded enough ($A_V$ greater than a few tenths) for molecules to form. That said, our adoption of a constant H$_2$ fraction is clearly an idealization. A better approach would be to self-consistently simulate the formation and destruction of H$_2$ in the numerical simulation assuming some illumination, so that the molecular fraction (as well as the ortho-to-para ratio) could be tracked. 

\begin{table*}
\begin{center}
\begin{tabular}{cccccccc}
\\
\hline \hline

 $N_{\rm{CH}^+}$ &  $N_{\rm{H}}$ & $\bar n_{\rm H}$ & $\sigma_v$ & $\bar T_{\rm M} $ & $\bar n_{\rm{H}, 99}$ & $\bar T_{\rm{M},99} $ & $\bar v_{d, 99}$ \\
$(10^{13}$ cm$^{-2}$) &$(10^{21}$ cm$^{-2}$) & (cm$^{-3}$) &(km s$^{-1}$) & (K) & (cm$^{-3}$) & (K) & (km s$^{-1}$)\\
\hline
1.1 & 2.2 & 36.0 & 1.8 & 60.0 & 1.5 & 713.0 & 2.1 \\
0.8 & 2.2 & 36.8 & 1.2 & 59.5 & 1.6 & 710.9 & 3.0 \\
1.1 & 1.2 & 19.8 & 1.9 & 92.5 & 1.7 & 708.5 & 2.3 \\
0.9 & 1.3 & 21.4 & 2.2 & 85.4 & 1.4 & 713.0 & 2.2 \\
0.6 & 1.9 & 31.2 & 2.2 & 69.3 & 1.9 & 682.3 & 1.7 \\
1.4 & 2.0 & 32.4 & 0.7 & 61.0 & 1.3 & 715.2 & 2.5 \\
1.6 & 1.9 & 31.7 & 1.8 & 66.9 & 1.9 & 676.6 & 2.2 \\
1.2 & 2.7 & 43.4 & 0.8 & 51.0 & 1.3 & 715.2 & 1.9 \\
\hline
\end{tabular}
\caption{Data from 8 randomly selected rays cast through the problem domain. \label{tbl-2} Column 1 - the CH$^+$ column density. Column 2 - the total column density. Column 3 - the mean number density. Column 4 - The 1D rms velocity dispersion. Column 5 - The mass-weighted mean temperature. Column 6 - The mean number density in the top 99\% of cells by CH$^+$ number density. Column 7 - The mass-weighted mean temperature in that same subset of cells. Column 8 - The mean drift velocity in the same cells. }
\end{center}
\end{table*}

\subsection{H$_2$ Emission}
\label{sec:H2}

\begin{figure}
  \centering
    \includegraphics[width=0.45\textwidth]{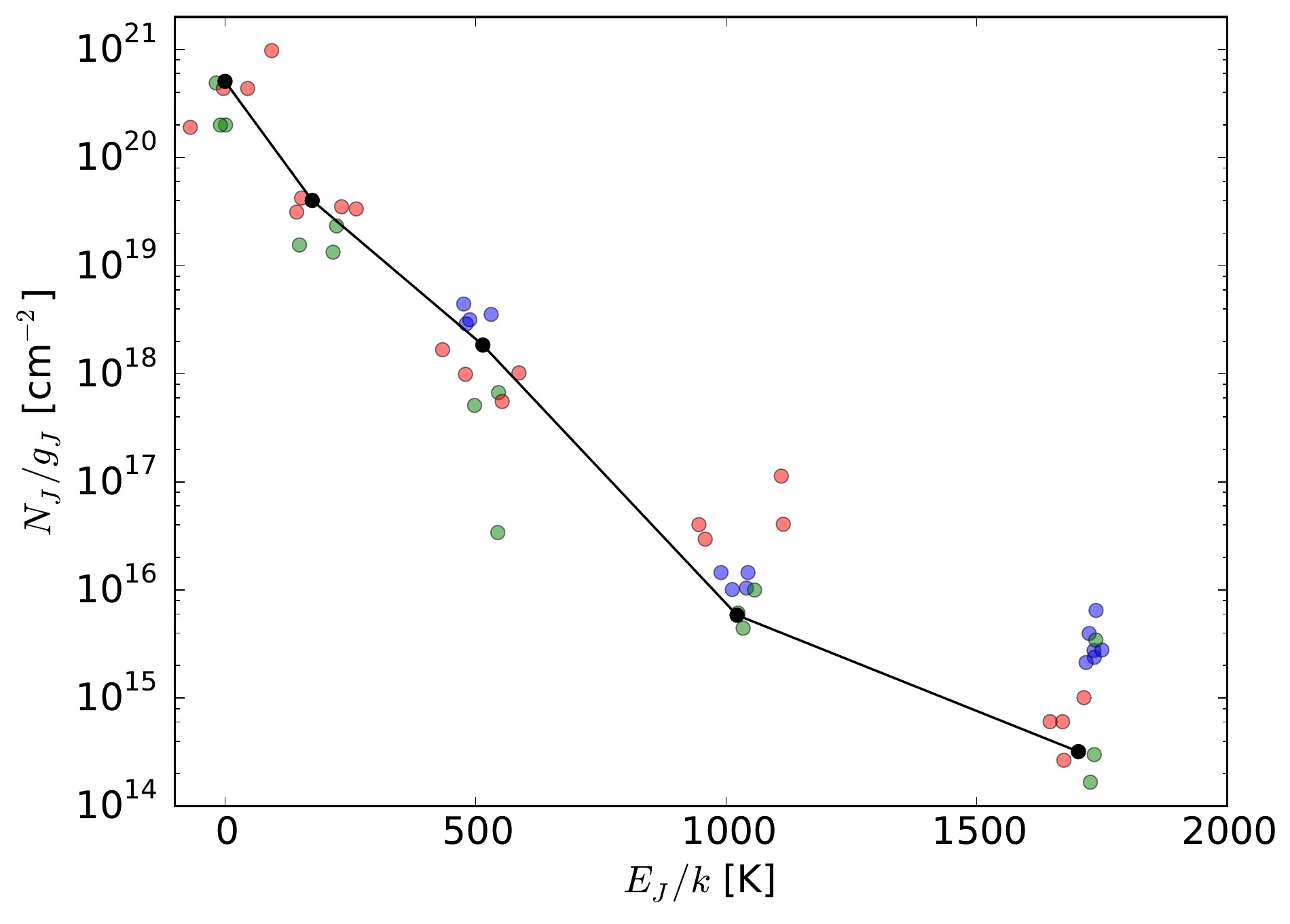}
    \caption{\label{fig:H2} Excitation diagram for the first 5 rotational levels of H$_2$ from our model compared to observed values. The $y$-axis shows the mean column densities in each rotational level divided by the degeneracy factors $g_J$. The $x$-axis is the energy of each level expressed as a temperature. Red circles - \protect\cite{lacour2005}. Green circles - \protect\cite{gry2002}. Blue circles - \protect\cite{ingalls2011}. Black circles - our model. We have scaled the column densities up by a factor of 3 to match the mean H$_2$ column density of the observations. The data have been shifted slightly along the $x$-axis for clarity.}
\end{figure}

At $T \gtrsim 1000$ K, significant numbers of H$_2$ molecules can be excited to $J \ge 2$ rotational levels, producing observable emission in the  $J=2 \rightarrow 0$, $J=3 \rightarrow 1$, and $J=4 \rightarrow 2$ lines. This emission is known to be correlated with CH$^+$ column density \citep{frisch1980, lambert86, jensen2010}, and has been interpreted as observational evidence for the intermittent dissipation of MHD turbulence \citep{falgarone2005}. Because diffuse clouds are typically below the critical densities of the $J = 3$ and higher lines, the level populations are in general non-thermal, and the level populations depend on both $T$ and the gas density. Observations of H$_2$ excitation can thus help to constrain the temperature and density structure in models of CH$^+$ production in diffuse clouds. To compare our results against these observations, we calculate the H$_2$ rotational level populations by balancing collisional excitation with collisional de-excitation and spontaneous emission for the first 20 rotational levels of H$_2$. We solve the resulting eigenvalue problem using SciPy's {\tt scipy.linalg.eig} routine. The full {\tt Cython}\footnote{http://cython.org/} code used for this calculation, as well as the rest of the code used in this paper, is available at \url{https://bitbucket.org/atmyers/chplus}.

Observations of $N_{\rm{H}_2}(J=0)$ and $N_{\rm{H}_2}(J=1)$ indicate that ortho- (odd $J$) and para- (even $J$) hydrogen are not generally in found in the equilibrium 3:1 ratio in the ISM. For example, in the three sight lines presented in \cite{gry2002}, $\gamma =  N_{\rm{H}_2}(J=1) / N_{\rm{H}_2}(J=0)$ is 0.7, 0.6, and 0.4. Likewise, in the four lines from \cite{lacour2005}, $\gamma = 0.3$, 1.6, 0.6, and 0.9. \cite{nehme2008} found $\gamma = 0.7$ towards HD 102065, and \cite{ingalls2011} found that this assumption fit their data for 6 nearby sight lines as well. In this paper, we will therefore fix the ortho:para ratio at 0.7 and treat ortho-H$_2$ and para-H$_2$ as separate species. For the energy levels $E_J$, we treat both forms of H$_2$ as quantum rotors with rotational temperature $T_r = 85.3$ K. We used the Einstein A coefficients from \cite{wolniewicz1998} and the collisional excitation rates from \cite{lebourlot1999}. The degeneracy factors are $g_J = (2 J+1)$ (para) and $g_J = 3 (2 J+1)$ (ortho). 

The result of this calculation for $J=0$ to $4$ is shown in Figure \ref{fig:H2}, compared against the H$_2$ excitation observations from \cite{gry2002}, \cite{lacour2005} and \cite{ingalls2011}. We have scaled all the column densities in our model up by a factor of 3 to match the mean H$_2$ column in the above papers. For all of the levels, our model result lies within the spread given by the observations. The mean $J = 1$ to $J=0$ rotational temperature in our calculation, defined by
\beq
\frac{N_{\rm{H}_2}(J=1)}{N_{\rm{H}_2}(J=0)} = \exp{ \frac{-171 \hspace{3 pt} \rm{K} } {T_{10}}},
\eeq is $\approx 67$ K, very close to the mean value for the sample of 38 sight lines in \cite{rachford2009} based on FUSE measurements. Because there is a wide range of temperatures present and because the $J = 2$ and higher lines are generally not in thermal equilibrium, however, our results for higher lines are not well-described by a single rotation temperature.

\section{Conclusions}

The intermittent dissipation of turbulence in the ISM has been proposed as an explanation for the high ($> 10^{13}$ cm$^{-2}$) CH$^+$ column densities commonly observed along diffuse molecular sight lines \citep{falg1995}. Turbulence can aid the production of CH$^+$ both by heating a small percentage of the gas directly \citep{lambert86, falg1995, pan2009} and by leading to large drift velocities \citep{spaans1995, fed1996, she2008} within localized regions of intense dissipation. Detailed dynamical and chemical models of these intense dissipation events have had much success in modeling the chemical properties of diffuse sight lines,
although the rate of strain in the dissipation events and their frequency in the ISM were assumed, not self-consistently calculated 
\citep{jou1998, god2009}.

We have re-assessed the origin of the CH$^+$ observed in the diffuse ISM by post-processing a direct numerical simulation of MHD turbulence, thereby self-consistently determining the properties of the intermittent turbulence that produces the CH$^+$.
We adopted the standard linewidth-size relation for molecular gas and set the total column density of our simulation equal to the mean value for the observed sample that we compare with; the resulting magnetic field in our simulation $\sim 5\;\mu$G is typical for interstellar gas with $\nbh\la 300$~cm\eee. We inferred that the density in our simulation is
$n_{\rm{H}} \simeq 30$ cm$^{-3}$, which implies that the size of the simulated region is $\ell_0 \approx 20$ pc and 
the velocity dispersion is $\sigma_{1\rm{D}} \approx 2.3$ km s$^{-1}$. Our approach provides the astrophysical framework in which the above models fit, and it corroborates their results in several ways. Specifically: 

\begin{enumerate}
\item We have solved an energy balance equation cell-by-cell for the temperature. While most of the mass in our cloud is cold ($\lesssim 35 $ K), a small fraction ($\lesssim 1 \%$) of the mass has been heated to temperatures in excess of 1000 K.
\item Similarly, we have computed the drift velocity in our simulation using an approximate analytic expression. While on average the drift is negligible, in isolated regions it can reach values equal to or exceeding the large scale RMS gas velocity of $\sim 4$ km s$^{-1}$ .
\item Both of these effects combine to easily produce CH$^+$ column densities in excess of $10^{13}$ cm$^{-2}$. We find that overall, the drift is more important, in that it alone accounts for about 2/3 of the CH$^+$ in the box, but the contribution from the gas temperature is not negligible. 
\item Our work highlights the importance of including density variations in physical and chemical models of the ISM. 90\% of the CH$^+$ is found in cells with densities of $\approx 4$ cm$^{-3}$ or less -- significantly lower than the mean value. These cells make up $\approx$ 5\% of the volume and $\approx$ 0.2\% of the mass in the simulation and have typical temperatures and drift velocities of $\approx 700 - 800$ K  and $\approx 3 - 4$ km s$^{-1}$. These values are quite similar to the TDR models of \cite{god2009}, despite the difference of our approaches.
\item We have estimated the CH$^+$ column density through our model and compared the resulting distribution to the sample of sight lines presented in \cite{wes2008}. Our mean CH$^+$ column of $1.3 \times 10^{13}$ cm$^{-2}$ agrees very well with the \citeauthor{wes2008} sample, although the median is lower by $\approx 26\%$.
\item Finally, we computed the expected H$_2$ rotational line emission from these hot regions, and found that it is consistent with observations of diffuse molecular sight lines.
\end{enumerate}

Finally, there are several caveats to our work that bear mentioning. The first, and probably most significant, is that our temperature and chemistry calculations were done purely in post-processing, taking the magnetic field, velocity, and density data from an isothermal MHD turbulent box. We cannot address whether or not our results would have been different had the temperature been computed self-consistently and allowed to affect the subsequent flow. We have also not self-consistently computed the H$_2$ fraction or the ortho-to-para ratio, instead using an average value inferred from observation. However, as chemical models of diffuse molecular gas frequently assume constant density, we believe that our approach is a valid first step towards a fully self-consistent turbulent chemistry calculation. A second, related caveat is that our results are tied to particular values of the sonic and \alfvenic Mach numbers that were used in the turbulence simulation. As discussed in Section 2.2, however, these dimensionless numbers correspond to reasonable choices for the cloud density ($\bar{n}_{\rm{H}} = 30 $ cm$^{-3}$), length scale ($L =$ 20 pc), and velocity dispersion ($\sigma_v = 2.3$ km s$^{-1}$). Furthermore, our choices for the Mach numbers yield results for the mean temperature, CH$^+$ abundance, and H$_2$ emission lines that are consistent with observations of diffuse molecular regions. A priori, the input parameters needed to ``predict" these observations could have been inconsistent with the typical properties of diffuse molecular gas, but we find that this is not the case. Third, we have used an approximate formula to estimate the ion-neutral drift velocity in the strongly-coupled regime (Section 2.4). While our formula has been calibrated against low-Mach number ($\mathcal{M} \approx 3$), non-ideal MHD simulations, we have extrapolated these results into the high-Mach number ($\mathcal{M} \approx 10$) regime, where they have not been directly verified. Doing so would require running non-ideal turbulence simulations with ambipolar diffusion in the high-Mach number, strongly coupled regime, which are not currently computationally feasible.  

\section*{Acknowledgments} 
ATM wishes to thank the anonymous referee for a thoughtful report that improved this paper. Support for this research was provided by NASA through NASA ATP grants NNX09AK31G and NNX13AB84G (CFM and PSL), the NSF through grants AST-0908553 and AST-1211729 (ATM and CFM), and the US Department of Energy at the Lawrence Livermore National Laboratory under grant LLNL-B569409 (A.T.M.).  This research was also supported by grants of high performance computing resources from the National Center of Supercomputing Application through grant TG-MCA00N020. We have used {\tt yt}\footnote{http://yt-project.org/}  \citep{turk2011} as well as the {\tt SciPy}\footnote{http://www.scipy.org/} family of Python libraries for data analysis and plotting. Our analysis and visualization scripts are available online at \url{https://bitbucket.org/atmyers/chplus}. 

\bibliographystyle{mn2e}
\bibliography{chplus}

\end{document}